\documentclass[prd,amsmath,amssymb,aps,floats,amsfonts,notitlepage,superscriptaddress,eqsecnum,nofootinbib,10pt]{revtex4-1}

\usepackage{booktabs}
\usepackage{graphicx}
\usepackage{dcolumn}
\usepackage{bm}
\usepackage{tensor}
\usepackage[utf8]{inputenc}
\usepackage{url}
\usepackage[colorlinks]{hyperref}
\usepackage{color}
\usepackage[dvipsnames]{xcolor}
\usepackage{mathtools}
\usepackage{tikz}
\usepackage{eufrak}
\usepackage[bbgreekl]{mathbbol}
\usepackage{comment}

\usepackage{enumitem}

\definecolor{CiteColor}{rgb}{0,0.5,0}
\hypersetup{citecolor=CiteColor}
\definecolor{RefColor}{rgb}{0.55,0,0}
\hypersetup{linkcolor=RefColor}
\definecolor{darkgreen}{rgb}{0.2,0.7,0.2}
\definecolor{cyan}{rgb}{0,0.9,0.9}
\definecolor{orange}{rgb}{0.9,0.5,0}
\definecolor{magenta}{rgb}{1,0,1}
\definecolor{purple}{rgb}{0.8,0.4,0.8}
\definecolor{gray}{rgb}{0.8242,0.8242,0.8242}
\definecolor{dodgerblue}{rgb}{0.12, 0.56, 1.0}
\definecolor{forestgreen}{rgb}{0.2, 0.6, 0.2}

\font\ec=ecrm0800 at 11pt
\def\th{\hbox{\ec\char'336}}
\def\edth{\hbox{\ec\char'360}}

\newcommand{\mb}{{\bar{m}}}

\newcommand{\ord}{\mathcal{O}}
\newcommand{\f}{\frac}

\newcommand{\pd}{\partial}
\newcommand{\be}{\begin{equation}}
\newcommand{\ee}{\end{equation}}
\newcommand{\nn}{\nonumber}
\newcommand{\s}{\sigma}
\newcommand{\flux}{\mathcal{F}}
\newcommand{\fluxI}{\mathcal{F}^\mathcal{I}}
\newcommand{\fluxH}{\mathcal{F}^\mathcal{H}}

\newcommand{\fluxbI}{\breve{\flux}^\mathcal{I}}
\newcommand{\fluxbH}{\breve{\flux}^\mathcal{H}}

\newcommand{\fluxbHPN}{\breve{\flux}^{\mathcal{H}(PN)}}
\renewcommand{\l}{\ell}

\newcommand{\p}{0}
\renewcommand{\O}{\Omega}

\newcommand{\m}{\mu}
\newcommand{\M}{M}
\newcommand{\hb}{\bar{h}}
\newcommand{\hbh}{\tilde{h}}
\newcommand{\E}{\mathcal{E}}

\newcommand{\geo}[1]{\hat{#1}}
\newcommand{\Og}{\geo{\Omega}}

\newcommand{\Os}{\Omega_\s}

\newcommand{\utsbar}{\breve{u}^t_\s}

\newcommand{\gencons}{\Xi}
\newcommand{\en}{\mathcal{E}}

\newcommand{\bpsi}{\Psi}
\newcommand{\ppsi}{\psi}
\newcommand{\hpsi}{\psi}

\begin{document}

\title{Dissipation in extreme-mass ratio binaries with a spinning secondary}

\author{Sarp Akcay}
\affiliation{Theoretisch-Physikalisches Institut, Friedrich-Schiller-Universität, 07743, Jena, Germany}
\affiliation{School of Mathematics \& Statistics, University College Dublin, Belfield, Dublin 4, Ireland}

\author{Sam R.~Dolan}
\affiliation{Consortium for Fundamental Physics, School of Mathematics and Statistics, University of Sheffield, Hicks Building, Hounsfield Road, Sheffield S3 7RH, United Kingdom}

\author{Chris Kavanagh}
\affiliation{Max Planck Institute for Gravitational Physics (Albert Einstein Institute), Am M\"{u}hlenberg 1, Potsdam 14476, Germany}
\affiliation{Institut des Hautes Etudes Scientifiques, F-91440 Bures-sur-Yvette, France}

\author{Jordan Moxon}
\affiliation{TAPIR, California Institute of Technology, Pasadena, CA 91125, USA}

\author{Niels Warburton}
\affiliation{School of Mathematics \& Statistics, University College Dublin, Belfield, Dublin 4, Ireland}

\author{Barry Wardell}
\affiliation{School of Mathematics \& Statistics, University College Dublin, Belfield, Dublin 4, Ireland}

\begin{abstract}
We present the gravitational-wave flux balance law in an extreme mass-ratio binary with a spinning secondary.
This law relates the flux of energy (angular momentum) radiated to null infinity and through the event horizon to the local change in the secondary's orbital energy (angular momentum) for generic (non-resonant) bound orbits in Kerr spacetime.
As an explicit example we compute these quantities for a spin-aligned body moving on a circular orbit around a Schwarzschild black hole.
We perform this calculation both analytically, via a high-order post-Newtonian expansion, and numerically in two different gauges. 
Using these results we demonstrate explicitly that our new balance law holds.
\end{abstract}
\maketitle

\section{Introduction}\label{Sec:Intro}

Gravitational wave physics is now firmly established as an observational science. Ground-based detectors regularly observe the binary mergers of stellar-mass black holes and neutron stars \cite{LIGOScientific:2018mvr}. 
Looking to the future, the construction of the space-based millihertz detector, LISA \cite{Audley:2017drz}, will open a new window on binaries with a total mass in the range $10^4$--$10^7 M_\odot$. 
One particularly interesting class of such systems are extreme mass-ratio inspirals (EMRIs) \cite{Babak:2017tow}.
In these binaries, a compact object, such as a stellar mass black hole or neutron star, spirals into a massive black hole driven by the emission of gravitational waves. 
These systems have a (small)  mass-ratio in the range of $10^{-4}-10^{-7}$.
In general EMRIs are not expected to completely circularize by the time of merger, resulting in a rich orbital and waveform structure that carries with it detailed information about the spacetime of the EMRI \cite{Amaro-Seoane:2014ela}. 
Additional complexity is added by the expectation that both the primary (larger) and secondary (smaller) compact object will be spinning, with no preferred alignment between the secondary's spin and the orbital angular momentum. 
Modelling the effects of the spin of the secondary is the focus of the present work.

Extracting EMRI signals from the LISA data stream will require precise theoretical waveform models of these binaries. 
This is because the instantaneous signal-to-noise ratio of a typical EMRI will be very small, and so the waveforms can only be separated from the instrumental noise and the potentially many other competing sources by semi-coherent matched filtering techniques \cite{Babak:2017tow}.

The small mass ratio in EMRIs naturally suggests black hole perturbation theory as a modelling approach. 
With this method, the spacetime of the binary is expanded around the analytically-known spacetime of the primary. 
The leading order contribution to the waveform phase comes from the orbit-averaged fluxes of gravitational radiation. 
These were calculated for a non-spinning secondary moving along a circular orbit about Kerr black hole in the 1970's \cite{Detweiler:1978ge}. 
These calculations were extended to eccentric \cite{Glampedakis:2002ya} and fully generic (inclined) motion \cite{Hughes:2005qb,Drasco:2005kz,Fujita:2009us} in the 2000's. 
The waveforms that can be constructed from these results will likely be sufficient for detection of the very loudest EMRIs. 
In order to detect the many weaker signals, to perform accurate parameter estimation, and to enable precision tests of general relativity, it is necessary to go beyond the leading-order model and include (so-called) \emph{post-adiabatic} contributions \cite{Hinderer:2008dm}.

The contributions at post-adiabatic order are substantially more challenging to calculate than the leading-order fluxes. 
This is because, often, the local metric perturbation near the secondary must be constructed and appropriately regularized whereas the leading-order fluxes can be computed from the asymptotic metric perturbation. 
Black hole perturbation calculations that involve the local metric perturbation are called \emph{self-force} calculations -- see  \cite{Poisson:2011nh, Wardell:2015kea, Barack:2018yvs} for reviews of foundations and calculation methods. 
With this in mind, the contributions to an EMRI waveform at post-adiabatic order come from the conservative and oscillatory dissipative first-order (in the mass-ratio) self-force, the orbit-averaged dissipative second-order self-force, and an orbit-averaged contribution from the spin of the small body. 
The first two of these have received a great deal of attention -- see Refs.~\cite{Barack:2007tm,Barack:2010tm,Akcay:2010dx,Akcay:2013wfa,Osburn:2014hoa,vandeMeent:2015lxa,vandeMeent:2017bcc} and \cite{Gralla:2012db,Pound:2012nt,Pound:2012dk,Warburton:2013lea,Pound:2014xva,Pound:2015wva,Wardell:2015ada,Miller:2016hjv,Pound:2017psq,Pound:2019lzj}, respectively. 
The influence of the secondary's spin on the inspiral has been less well studied and is the topic of the present work.

Our goal here is to understand how the inspiral (and by extension the waveform from the EMRI) is influenced by the spin of the secondary. 
For a non-spinning secondary, well-known balance laws \cite{Mino:2003yg, Sago:2005gd, Sago:2005fn, Hughes:2005qb, Drasco:2005is, Ganz:2007rf, Isoyama:2018sib} can be used to relate the leading-order fluxes to the first-order self-force contribution to the evolution of the inspiral. 
In this work we derive, for the first time, the appropriate balance law including the contribution from the spin of the secondary in the extreme mass-ratio inspiral context.

We obtain the flux balance law: For a small companion with spin, the flux of
energy $\flux = \fluxI + \fluxH$ out to future null
infinity, $\mathcal{I}^+$, and down the horizon, $\mathcal{H}^+$, (which can be evaluated entirely from metric
perturbation $h_{\mu \nu}$ at $\mathcal{I}^+$ and $\mathcal{H}^+$)  is equivalent to the
rate of change of the quasi-conserved energy $\mathcal{E}$ associated with the
spin and orbital motion of the small companion to linear order in the mass ratio
and spin of the small companion.
For the case of quasicircular orbits, we obtain the succinct result
  \begin{equation} \label{eq:SummaryEquation}
    \frac{D\mathcal{E}}{d \tau}
    = \frac{1}{2} u^\alpha u^\beta \mathcal{L}_\xi h_{\alpha \beta}^{\mathcal{R}}
    - \frac{1}{2 \mu} S^{\gamma \delta} u^\beta \nabla_\delta \mathcal{L}_\xi h^{\mathcal{R}}_{\gamma \beta}
    = u^t \flux,
  \end{equation}
   where $\xi$ is the timelike Killing vector of the background metric, $\mu$ and
  $S^{\alpha \beta}$ are the mass and the spin tensor of the small companion.
  $h^{\mathcal{R}}_{\mu \nu}$ is the Detweiler-Whiting regular part of the
  metric perturbation, and $u^\alpha$ is the worldline four-velocity.
 Further, we derive that an orbit-averaged version of
  \eqref{eq:SummaryEquation} holds for generic orbits and arbitrary Killing
  vector in Kerr spacetime.
 In particular, our result is directly applicable to the quasiconserved angular
  momentum $L_z$ associated with the angular Killing vector of black hole
  backgrounds.
 The intermediate geometric result of \eqref{eq:SummaryEquation} can be
  obtained either from direct expansion of traditional self-force formulas (as
  we show in Section~\ref{sec:GeoIdentities}), or from a specialization and multipole expansion
  of results from Ref.~\cite{Harte:2014wya}.

The power of obtaining such flux balance laws are twofold. First, providing a direct relation between the local metric perturbation and the asymptotic losses of energy and angular momentum gives a gauge-invariant tool for checking the dissipative part of local self-force calculations. 
In the non-spinning case these have long been used for benchmarking \cite{Barack:2010tm,vandeMeent:2017bcc}. 
Second, flux balance laws enable a dramatic simplification of the computational cost in computing the effects of the orbit averaged dissipative self-force; fluxes are much easier to compute than local self-forces as they only require knowledge of the asymptotic and not the local metric perturbation. 
In the non-spinning case the net result of this statement is that to adiabatic order, the fluxes are entirely sufficient to drive an inspiral. 
For the case of a spinning secondary which we consider here, the situation is more complicated. 
The fluxes will be sufficient in determining the evolution of constants of motion associated with Killing vectors $\xi_\alpha$. 
However, these constants of motion $\gencons$ will determine the four-velocity $u^\alpha$ using the following relation:
\begin{equation}
  \gencons = u^\alpha \xi_\alpha
  + \frac{1}{2 \mu} S^{\alpha \beta} \nabla_\alpha \xi_\beta.
\end{equation}
Thus, to determine the evolution of the 4-velocity one will need also to evolve the spin tensor $S^{\alpha \beta}$. 
The governing equation for this evolution will be given in Sec.~\ref{Sec:Motion} and requires knowledge of the local metric perturbation.

We explicitly verify our balance law in the case of a spinning body whose spin vector is aligned with the orbital angular momentum, and which is moving along a circular orbit of a Schwarzschild black hole. We perform this calculation in two gauges: the radiation gauge (via the Teukolsky formalism) and Lorenz gauge. In the former approach, we made our computations both numerically and analytically
(as a high-order post-Newtonian expansion) and in the latter approach the computations were carried out numerically.
We find excellent agreement between the two gauges for the (gauge-invariant) fluxes and local dissipative force. 
We also confirm that our flux balance law \eqref{eq:SummaryEquation} holds to the numerical precision of our calculation, and exactly (to the relevant PN order) in the analytic case.

It is important to note that our work is not the first calculation of the radiated flux, $\flux$, for a spinning body. 
These have been carried out before \cite{Han:2010tp, Harms:2015ixa, Harms:2016ctx, Lukes-Gerakopoulos:2017vkj, Nagar:2019wrt} (though we perform our calculations to a much higher precision). 
Our work presents, for the first time, the derivation of a new balance law for spinning bodies; the first calculation of the local dissipative force; an explicit numerical check that this balance law holds; and a comparison with a post-Newtonian expansion at $5.5$pN order.

The layout of this paper is as follows. 
In Sec.~\ref{Sec:Motion} we provide the self-forced equations of motion for a spinning body. 
In Sec.~\ref{sec:GeoIdentities} we derive the balance law including the contribution from the spin of the secondary. 
This calculation is valid for arbitrary non-resonant orbital configurations to linear order in the spin of the secondary. 
In Sec.~\ref{Sec:DipoleSource} we specialize to the case of a spin-aligned body on a circular orbit about a Schwarzschild black hole.
In Sec.~\ref{sec:TeukolskyCalculation} we describe the calculation of the fluxes and local force within the Teukolsky framework (with further details given in the Appendices). 
In Sec.~\ref{Sec:LorenzGauge} we do the same, but in the Lorenz gauge. 
The results of these two sections are compared in Sec.~\ref{Sec:Results} and we conclude with Sec.~\ref{Sec:Conclusions}. 
Throughout this work we used geometrized units such that the speed of light and the gravitational constant are set to unity ($G=c=1$). We define $M$ to be mass of the primary.
We use both prefix ($\nabla_\alpha$) and postfix ${}_{;\alpha}$ notations for covariant derivatives, choosing the notation that is most clear in a given expression. We denote symmetrization of indices using round brackets [e.g. $T_{(\alpha \beta)} = \tfrac12 (T_{\alpha \beta}+T_{\beta \alpha})$] for symmetrization and square brackets [e.g. $T_{[\alpha \beta]} = \tfrac12 (T_{\alpha \beta}-T_{\beta \alpha})$] for antisymmetrization, and exclude indices from symmetrization by surrounding them by vertical bars [e.g. $T_{(\alpha | \beta | \gamma)} = \tfrac12 (T_{\alpha \beta \gamma}+T_{\gamma \beta \alpha})$].

\section{Self-forced equations of motion for a spinning companion} \label{Sec:Motion}

We consider an object of mass $\m$ in a binary system with a black hole of much
greater mass $\M \gg \m$.
Both companions are permitted to possess spin, which we denote in scaling
arguments as $S_1$ for the spin of the primary and $S_2$ for the spin of the
secondary.
The perturbative expansion of the equations of motion and field equations are
performed using the mass ratio $\epsilon \equiv \m / \M$ and the dimensionless spin
parameter $\s \equiv S_2 / \m \M$ (henceforth we refer to $\s$ as the ``spin''
of the secondary).
We consider self-force effects to linear order in  $\s$ and $\epsilon$.
Concretely, a system for which this expansion is relevant is one for which
$\epsilon \ll \s \ll 1$, and for which higher multipole moments contribute at
$\mathcal{O}(\s \epsilon^2)$.

In fact, the analysis and balance law which we present are perfectly valid for
determining the contributions linear in spin in the more generic case where
$\epsilon \ll 1$ and $\sigma \ll 1$ hold separately.
For the generic case, the analysis presented here does not give a complete
approximation for the equations of motion, as other effects will enter at
orders comparable to the linear-in-spin contributions presented below.
However, the linear-in-spin effects are fully captured by our analysis, so a
complete perturbation can be obtained by simply adding the
$\mathcal{O}(\sigma \epsilon)$ part described here to spin-independent
contributions at the same perturbative order.

The most relevant case for self-force computations is $\epsilon \sim \sigma$,
which describes a compact secondary, such as a black hole or neutron star.
For this case, the leading spin effects will enter at $\mathcal{O}(\epsilon^2)$,
which is the same order as the second-order self-force.
Therefore, for spinning bodies, the leading spin contribution discussed here
should be regarded as similarly important for full phase accuracy as the
second order self-force pursued by other investigations
\cite{Pound:2019lzj,Pound:2017psq}.

We consider a perturbation of the form
\begin{equation}
  \mathbf{g}_{\alpha \beta}  = g_{\alpha \beta} + h_{\alpha \beta} + \mathcal{O}(\epsilon^2),
\end{equation}
where our goal is to capture in $h_{\alpha \beta}$ the contributions from the small companion through $\mathcal{O}(\sigma)$. We neglect effects which are second-order in the mass
ratio, quadratic and higher in the spin of the small companion, or of quadrupole or higher
multipole order. For brevity, we use the notation $\mathcal{O}(\epsilon^2)$ to indicate
that we are neglecting all of these higher-order contributions.

The fully general form for the self-force on an extended body may be derived
from a Green's function treatment of the metric perturbation sourced by that
body.
A careful presentation of the generic spacetime integrals required to derive the
self-force equations of motion to arbitrary order in the mass ratio and to
arbitrary multipolar order were derived and extended by Refs.~\cite{Dixon:1970zza,Dixon:1970zz,Dixon-3,Harte:2011ku,Harte:2014wya}.
We refer to the set of equations obtained by the derivation in those
publications as the Dixon-Harte equations of motion.
In this work, we make use of the Dixon-Harte equations of motion specialized to
first order in the mass ratio $\epsilon$ and dimensionless spin $\sigma$.
Below, we describe these specializations first to leading (zeroth) order in
self-field effects, obtaining the Matthisson-Papapetrou-Dixon equations for a
freely falling point particle with spin in an arbitrary background spacetime; next,
we show the specialization for the less well-known equations of motion to linear
order in the spin and self-field effects.

\subsection{Perturbative expansion of the self-forced motion} \label{sec:PerturbedEOM}

The Dixon-Harte formalism derives the equation of motion for an extended body
(in our case, the small companion) under the effects both of the background
metric associated with the large companion $g_{\alpha \beta}$, and of the metric
perturbations sourced by the secondary's own motion.
The generic result is the evolution equations of the overall momentum and spin
of the small object in terms of linear combinations of four-integrals over the
stress-energy distribution of the body.
Due to the length of the expressions and their notational complexity, we do not
reproduce the generic expressions here, and instead refer the interested reader
to their full presentation \cite{Dixon-3,Harte:2014wya}.
Wherever possible, we follow the notation of Ref.~\cite{Harte:2014wya}, and note below all exceptions where we specialize or
deviate from that notation.

For the present discussion, we make use of the linear momentum vector $p^\mu$
and the spin tensor $S^{\alpha \beta}$, defined along the center-of-mass worldline of
the small companion.
We define these quantities on a choice of hypersurface foliation $\Sigma$, and
with respect to a worldline $z^\mu(\tau)$ for proper time $\tau$ along that
worldline.
Note that the generic treatment by Ref.~\cite{Harte:2014wya} uses the distinct time
variable $s$, which reduces to $\tau + \mathcal{O}(\epsilon^2)$ under the
specializations used in this paper.

We use Synge's worldfunction $\sigma(z^\mu, x^{\mu^\prime})$ \cite{Synge:1960}
and its
derivatives for a covariant notion of distance and displacement vectors. (Notationally, Synge's worldfunction is here distinguished from the spin parameter $\sigma$ by its bitensor arguments.) 
Synge's worldfunction is a bitensor which takes the value of half the square of
the affine parameter $\lambda^2 /2$ of the geodesic which joins the
points $z^\mu$ and $x^{\mu^\prime}$. 
The first covariant derivative of Synge's worldfunction
$\sigma_{; \mu^\prime}(z^\mu, x^{\mu^\prime}) \equiv \sigma_{\mu^\prime}(z^\mu,
x^{\mu^\prime})$ is a covariant analog of the displacement vector between the
two points, in the tangent space of $x^{\mu^\prime}$.
Further details regarding bitensors and Synge's worldfunction may be found in
Ref.~\cite{Poisson:2011nh}.
In particular, the relationship between the tangent vector, unique shortest
geodesic between the two spacetime points, and Synge's worldfunction is nicely
illustrated in Fig.~5 of Ref.~\cite{Poisson:2011nh}.
The full Dixon-Harte formalism proceeds using an intricate bitensor treatment
necessary for a nonperturbative description of linear and angular momentum
evolution.
For the perturbative expansion in powers of the mass ratio $\epsilon$, the linear momentum and spin of the small
companion to the order required by this paper are
\begin{subequations} \label{eq:MomentumSpinDef}
  \begin{align}
    p^\mu &= \int d \Sigma_{\nu^\prime} T^{\nu^\prime \mu^\prime}(x^\prime)\,
            g^\mu{}_{\mu^\prime}(z^\alpha, x^{\alpha^\prime}), \\
    S^{\mu \nu} &= \int d \Sigma_{\nu^\prime} T^{\nu^\prime \mu^\prime}(x^\prime)\,
                  g^{[\mu}{}_{\mu^\prime}(z^\alpha, x^{\alpha^\prime})
                  \,\sigma^{\nu]}(z^\beta, x^{\beta^\prime}),
  \end{align}
\end{subequations}
in which the primed indices are used for the tangent space away from the
worldline, and $g^\mu{}_{\mu^\prime}$ denotes the parallel propagator.
The rest mass of the small companion is related to the linear momentum vector by
$\mu = \sqrt{-p_\mu p^\mu}$.

In the Dixon-Harte construction, the center-of-mass worldline is freely
specifiable in the definition of the multipole moments; different choices of
$z^\mu$ give rise to different values of $p^\mu$ and $S^{\alpha \beta}$ while
preserving the form of the resulting equations of motion.
To fix the remaining freedom in $z^\mu$, one makes a choice of the
center-of-mass condition, often choosing components of the spin tensor
$S^{\alpha \beta}$ to be considered as the `mass dipole' and setting those
components to zero.
This type of constraint on the spin tensor is referred to as a `spin
supplementary condition'.
For this paper, we work with moments defined using the Tulczyjew spin
supplementary condition \cite{Tulczyjew:1959}
\begin{equation}
  S^{\alpha \beta} p_{\beta} = 0.
\end{equation}

Applying the expansion in powers of small separation from the worldline
$\sigma(z^\mu, x^{\mu\prime}) \ll M^2$ and in powers of the dimensionless parameters
$\epsilon$ and $\sigma \epsilon$ that parameterize the strength of the metric
perturbation sourced by the small companion, the leading order motion derived by
specializing the Dixon-Harte formalism reduces to the well-known
Mathisson-Papapetrou-Dixon (MPD) equations of motion for a spinning test
particle \cite{Mathisson:1937zz,Papapetrou:1951pa,Dixon:1970zza}
\begin{subequations} \label{eq:MPD}
\begin{align}
 	\frac{Dp^\alpha}{d\tau}
  &= -\frac{1}{2}\tensor{R}{^\alpha_\beta_\gamma_\delta}
    u^\beta S^{\gamma\delta} + \mathcal{O}(\epsilon)	,\\
	\frac{DS^{\gamma\delta}}{d\tau}
  &= 2 p^{[\gamma}u^{\delta]}+ \mathcal{O}(\epsilon).
\end{align}
\end{subequations}

The Dixon-Harte formalism also offers a prescription for determining the
expansion of $p^\mu$ in terms of the worldline four-velocity
$u^\mu \equiv \frac{D z^\mu}{d\tau}$ and the higher multipole moments of the small
companion.
Performing the specialization to the present perturbative treatment, we find
that this relationship is simply
\begin{equation}
  p^\alpha = \m u^\alpha + \mathcal{O}(\epsilon^2),
\end{equation}
and therefore, the leading equations of motion may also be written as
\begin{subequations}
  \begin{align}
    a^\alpha \equiv \frac{D u^\alpha}{d\tau}
    &= - \frac{1}{2 \m}
      R^\alpha{}_{\beta \gamma \delta} u^\beta S^{\gamma \delta}
      + \mathcal{O}(\epsilon), \\
    \frac{D S^{\gamma \delta}}{d\tau} &= \mathcal{O}(\epsilon).
  \end{align}
\end{subequations}

For use in subsequent sections, it is also useful to invert the Dixon-Harte
moments \eqref{eq:MomentumSpinDef} expanded in the mass ratio $\epsilon$ for the
monopole and dipole moments to obtain a series expansion for the stress energy
tensor
\begin{equation}\label{eq:Texpansion}
  T_{\alpha \beta} = \m T^{(\m)}_{\alpha \beta} + \m \s T^{(\s)}_{\alpha \beta}
  + \mathcal{O}(\epsilon^2),
\end{equation}
where both $T^{(\m)}_{\alpha\beta}$ and $T^{(\s)}_{\alpha\beta}$ are $\mathcal{O}(1)$ (note, however, that they both have subleading dependence on $\s$ and $\epsilon$ via the worldline).
To simplify the expression of $T_{\alpha \beta}^{(\sigma)}$, we introduce the
scaled spin parameter $\tilde{S}^{\mu \nu} \equiv S^{\mu \nu}/(\sigma \mu)\sim \mathcal{O}(M)$. 
Then, the two contributions to the stress-energy are
\begin{subequations}
\begin{align}
    T^{(\m) \alpha \beta}(x)
      &= \int d\tau \frac{\delta^{4}(x^\mu -
        z^\mu(\tau))}{\sqrt{-g}}  u^\alpha(\tau) u^\beta(\tau),\\
    T^{(\s) \alpha \beta}(x)
      &= \int d\tau \nabla_\delta \left(\frac{\delta^4(x^\mu -
        z^\mu(\tau))}{\sqrt{-g}} \right)
        u^{(\alpha}(\tau) \tilde{S}^{\beta) \delta}(\tau).
\end{align}		\label{eq:Tmunu}
\end{subequations}

Our use of the Dixon-Harte formalism is primarily motivated by the requirement
of having a rigorous foundation for the next order of perturbative expansion
which contains the first order monopole-sourced self-force, the first order
spin-sourced self-force, and the first-order self-torque.
All of these ingredients prove important in the full flux balance law for a
small companion with spin, as shown in Sec.~\ref{sec:GeoIdentities}.

Using the fact that perturbations to the connection and the Riemann tensor can be written as tensor expressions with respect to the background $g_{\alpha\beta}$, given 
by\footnote{Here, as in the rest of the paper we omit for notational compactness the explicit dependence of expanded quantities on $g_{\alpha \beta}$, and simply note that the Riemann tensor and covariant derivative on the right hand side are those associated with the background $g_{\alpha \beta}$.}
\begin{equation}
  R_{\alpha\beta\gamma\delta} (\mathbf{g}) = R_{\alpha\beta\gamma\delta}
  + \left(h_{\alpha}{}^{\lambda} R_{\lambda \beta \gamma \delta}
   - h_{\beta [\delta; |\alpha| \gamma]}
   + h_{\alpha [\delta; |\beta| \gamma]}
   + h_{\alpha \beta; [\delta \gamma]}\right)\epsilon  + \mathcal{O}(\epsilon^2),
   \label{eq:deltaR}
\end{equation}
\begin{equation}
 \Gamma^{\alpha}_{\beta \gamma} (\mathbf{g}) - \Gamma^{\alpha}_{\beta\gamma}(g) = \tfrac12 g^{\alpha \delta} \left(h_{\beta \delta;\gamma} + h_{\delta \gamma; \beta} - h_{\beta\gamma;\delta}\right)\epsilon  + \mathcal{O}(\epsilon^2)
\end{equation}
and also accounting for the perturbation to the proper time (see, e.g., Sec.~19.1 of Ref.~\cite{Poisson:2011nh}),
we now expand the Dixon-Harte equations of motion to subleading order in the mass
ratio $\epsilon$ and the dimensionless spin parameter $\sigma$.
Making use of the notation common in the self-force literature which
constructs a separation between the `singular' and `regular' parts of the metric
perturbation, and
denoting the regular part of the metric perturbation with superscript $\mathcal{R}$, we
find the equations of motion through $\mathcal{O}(\epsilon)$ are

\begin{subequations} \label{eq:LeadingSpinSelfForce}
  \begin{align} \label{eq:SpinAcceleration}
    a^{\alpha}
    =& \frac{1}{2} \sigma u^\beta \tilde{S}^{\delta \epsilon} R_{\lambda \beta \epsilon \delta} - \frac{1}{2} \epsilon \left(g^{\alpha \lambda} + u^\alpha u^\lambda\right)
       \bigg[u^\gamma u^\delta \left(2 h^{\mathcal{R}}_{\gamma \lambda; \delta}
       -  h_{\gamma \delta; \lambda}^{\mathcal{R}}\right) -\sigma u^\beta \tilde{S}^{\delta \epsilon}
       \left(
       h_{\lambda\gamma}^{\mathcal{R}} R^\gamma{}_{\beta \epsilon \delta}
       - h^{\mathcal{R}}_{\beta \delta; \lambda \epsilon}
       + h^{\mathcal{R}}_{\lambda \delta;\beta \epsilon}
       + h^{\mathcal{R}}_{\beta \lambda; \delta \epsilon} \right) \bigg],
       \\
       \frac{D \tilde{S}^{ \gamma \delta}}{d \tau}
    =& - \epsilon \sigma  u^\alpha \tilde{S}^{\beta [\delta} g^{\gamma] \lambda}
       \left( h^{\mathcal{R}}_{\lambda \beta; \alpha}
       + h^{\mathcal{R}}_{\alpha \lambda; \beta}
       - h^{\mathcal{R}}_{\alpha \beta; \lambda}\right) - \frac{1}{2} \epsilon \sigma \tilde{S}^{\gamma \delta} u^\alpha u^\beta u^\lambda h^{\mathcal{R}}_{\alpha \beta;\lambda},
       \label{eq:SpinTorque}
  \end{align}
\end{subequations}
where $h_{\alpha \beta}^{\mathcal{R}}$is the Detweiler-Whiting regular field.
In the limit $\sigma \to 0$, the first of these equations becomes the well-known MiSaTaQuWa
(self-force) equation of motion \cite{Mino:1996nk,Quinn:1996am}.

\section{Flux-balance law to linear order in spin}
\label{sec:GeoIdentities}

The perturbative context in which we work leads us to a description of the
motion of the small companion and the radiation that it sources as perturbed
fields in the background spacetime of the large companion.
Then, any symmetries of the background spacetime might be anticipated to give
rise to conservation laws, such that certain quantities near the small companion
might be inferred from field data far from the system.
The relevant symmetries can be described using the Killing vectors of the
background spacetime, which obey the defining property
\begin{equation}
  \nabla_{(\alpha} \xi_{\beta)} = 0.
\end{equation}
Specifically, for Schwarzschild and Kerr spacetimes, there exist two Killing
vectors $\xi_t^\mu = \{1, 0, 0, 0\}$ and $\xi_\phi^\mu = \{0, 0, 0, 1\}$,
associated with the invariance of the metric under time translations and
rotations.
We note that the Killing tensor of the Kerr metric should also be anticipated to
give a balance law, associated with a relationship between the Carter constant
of the small companion's orbit and asymptotic field quantities, but this
derivation for a spinning body is left for future explorations of the topic.

\begin{figure} 
  \includegraphics[width=.55\textwidth]{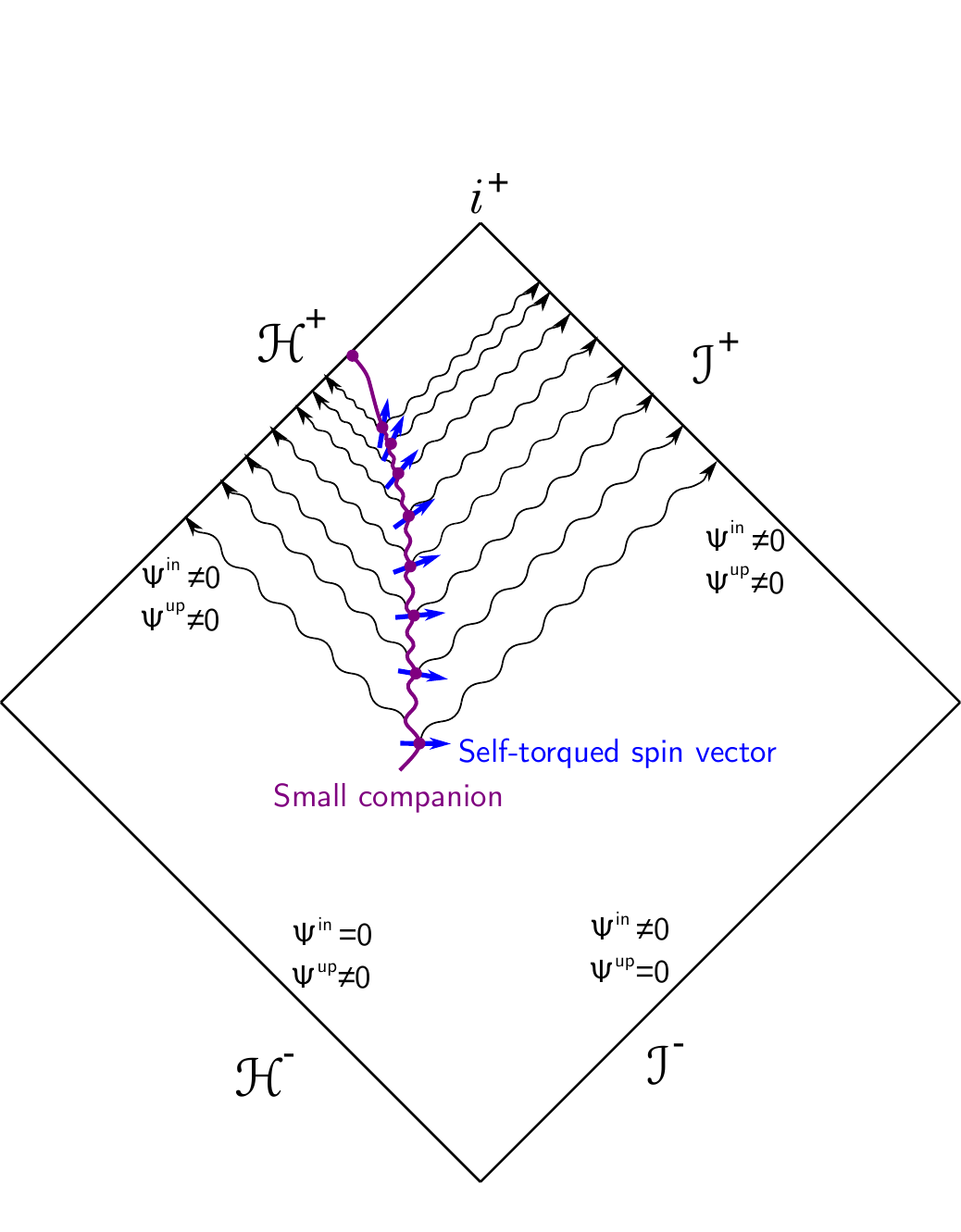}
  \caption{As the small companion's energy and angular momentum from its orbit
    and spin vector evolves, generating gravitational radiation, an equivalent
    orbit-averaged flux of energy and angular momentum escapes to
    $\mathcal{I}^+$ and down the horizon $\mathcal{H}^+$ . The ``in'' and ``up''
    modes used in this paper are constructed to vanish at
    $\mathcal{I}^-$ and $\mathcal{H}^-$, respectively. Both sets of modes have nonvanishing
    contribution to the flux at $\mathcal{H}^+$  and $\mathcal{I}^+$.
    Note that the arrow indicating the spin vector is shown as a cartoon of the
    secular spin evolution, and the orientation does not have detailed meaning
    with respect to the axes of the spacetime diagram.}
    \label{fig:spacetime_diagram}
\end{figure}

It is a well-known property that geodesic orbits of pure monopole masses 
($S^{\alpha \beta}$ = 0) preserve the orbital parameters
\begin{equation}
\gencons^{(\mu)} = u^\alpha \xi_\alpha,
\end{equation}
for each Killing vector $\xi$ of the background spacetime.
The two conserved parameters obtained from Schwarzschild Killing vectors
are interpreted as the energy and angular momentum for the timelike and 
angular Killing vectors, respectively.

The conservation law for general test-mass (for which $S^{\alpha \beta} \ne 0$)
motion follows similarly, and gives rise to the result that the MPD equations of
motion (\ref{eq:MPD}) preserve the conserved parameters
\begin{equation}
  \gencons = u^\alpha \xi_\alpha
  + \frac{\sigma}{2} \tilde{S}^{\alpha \beta} \nabla_\alpha \xi_\beta. \label{eq:cons_E}
\end{equation}
These adjusted conserved parameters have the interpretation of the sum of energy
and angular momentum contributions from the orbital motion of the small
companion and from its intrinsic spin.
The constancy of these parameters gives rise to important simplifications in the
derivation of the test mass motion, and the extension of such identities to
radiation-reaction motion offers the possibility of computing aspects of the
adiabatic evolution of self-forced orbits from field variables.
Such a computation can then avoid potentially costly local computation of the
instantaneous force on the small companion.

``Flux-balance'' laws for EMRI motion are concrete conservation identities
between the evolution of the now quasi-conserved quantities (\ref{eq:cons_E}) of
the small companion during radiation-reaction and quantities computable from
gravitational wave amplitudes evaluated at the null surfaces of the future
horizon $\mathcal{H}^+$  and future null infinity $\mathcal{I}^+$ for a black hole
inspiral in an asymptotically flat spacetime.
Despite the difficulty in defining a reasonable effective energy or angular
momentum associated with the gravitational perturbations sourced by the small
object within the strong-field region, these balance laws offer the simple
interpretation of an amount of energy or angular momentum ``lost'' to radiation
by the small companion, and escaping to $\mathcal{H}^+$  or $\mathcal{I}^+$ in the
form of gravitational waves (see Fig \ref{fig:spacetime_diagram}).

There are several existing derivations which show the direct relation between
the evolution of the energy, angular momentum, and Carter constant (for Kerr
backgrounds) for monopolar test masses.
The original flux-balance law derivation by Ref.~\cite{Galtsov:1982hwm} made use of
the evolution of the momentum of the small object directly in terms of radiative
fields, which were then used in a calculation using the Green's function for the
fields to show the balance of energy and angular momentum.
The derivation was subsequently extended to the Carter constant by
Ref.~\cite{Mino:2003yg}.
In Ref.~\cite{Sago:2005fn}, the flux balance law for the energy, angular momentum,
and Carter constant was derived using a simpler mathematical method, which
serves as the foundation for our derivation.
Due to the similarity of the methods for the monopole and dipole computations,
we anticipate that the flux-balance law for a small companion with spin could
also be extended to the Carter constant.
A more recent investigation \cite{Isoyama:2018sib} has applied a Hamiltonian
method to extend flux-balance relations to resonant orbits.

\subsection{Conservation identity including spin}
\label{sec:GeoIdentityMonopole}

Each Killing vector, $\xi^\mu$, of a spacetime gives rise to a conserved orbital
parameter, $\gencons$, for test-body motion in that spacetime.
Taking advantage of the defining property of a Killing vector
$\nabla_{(\alpha} \xi_{\beta)} = 0$, we derive the equation of motion for the
conserved quantity $\gencons$ by differentiating Eq.~\eqref{eq:cons_E},
\begin{equation} \label{eq:fluxeqstart}
  \frac{D \gencons}{d\tau} \equiv u^\alpha \nabla_\alpha \gencons
  = \xi^\beta a_\beta
  + \frac{\s}{2} u^\alpha \nabla_\alpha
  \left( \tilde{S}^{\gamma \beta} \right) \nabla_\gamma \xi_\beta
  + \frac{\s}{2} u^\alpha \tilde{S}^{\gamma \beta} \nabla_\alpha \nabla_\gamma \xi_\beta.
\end{equation}

We are interested primarily in the overall dissipation of the orbital conserved
quantities, and wish to ignore in these computations any oscillatory changes
that might occur during the interaction of the small companion and its
radiation.
To evaluate the dissipative effects as separate from any conservative
oscillations, we define an ``orbital'' averaging operation $\langle \dots \rangle$:
\begin{equation} \label{eq:crudeAA}
  \langle f(\tau)\rangle \equiv \frac{1}{2T} \int_{\tau-T}^{\tau + T} d\tau^\prime
    f(\tau^\prime) ,
\end{equation}
where the limits of the integral are understood to obey the
restriction $\M^2/\m \gg T \gg \M$.
A more careful formulation of an orbit averaging operator can be obtained by use
of multiscale techniques \cite{Hinderer:2008dm, Pound:2015wva}.
The expression (\ref{eq:crudeAA}) is not purely an average over oscillatory
degrees of freedom, as the radiation-reaction force will cause the orbital
parameters to evolve slightly over time $T$.
However, neglecting resonances, the difference in the above orbit-average and a
version which treats the oscillatory contributions more carefully is
second order in the mass ratio $\epsilon$, so may be neglected in our
derivation.

We now consider the expansion of equation
\eqref{eq:fluxeqstart} for the evolution of the conserved quantity
$\gencons$ associated with the Killing vector $\xi^\mu$.
First, we expand \eqref{eq:fluxeqstart} by substituting the
acceleration \eqref{eq:SpinAcceleration} and torque \eqref{eq:SpinTorque}.
In addition, it is useful to apply the identity for the second covariant
derivative of a Killing vector
\begin{equation} \label{eq:KillingIdent}
  \nabla_\alpha \nabla_\gamma \xi_\beta
  = \xi^\delta R_{\delta \alpha \gamma\beta}.
\end{equation}
Combining the contributions to the orbit-averaged flux value we find
\begin{align}
  \left\langle \frac{D \gencons}{d \tau} \right\rangle =
  \bigg\langle
  &- \frac{1}{2} \xi^\alpha (g_\alpha{}^\lambda +  u_\alpha u^\lambda)
    \bigg[u^\gamma u^\delta \left(2h^{\mathcal{R}}_{\gamma \lambda; \delta}
    - h^{\mathcal{R}}_{\gamma \delta; \lambda}\right) -\sigma  u^\beta \tilde{S}^{\delta \epsilon}
       \left(
       h^{\mathcal{R}}_{\lambda\gamma} R^\gamma{}_{\beta \epsilon \delta}
       - h^{\mathcal{R}}_{\beta \delta; \lambda; \epsilon}
       + h^{\mathcal{R}}_{\lambda \delta;\beta; \epsilon}
       + h^{\mathcal{R}}_{\beta \lambda; \delta; \epsilon} \right) \bigg]\notag\\
  & -\frac{1}{2} u^\alpha \sigma \tilde{S}^{\beta \delta} g^{\gamma \lambda}
    \left( h^{\mathcal{R}}_{\lambda \beta; \alpha} + h^{\mathcal{R}}_{\alpha \lambda; \beta}
    - h^{\mathcal{R}}_{\alpha \beta; \lambda}\right)  \xi_{\delta;\gamma} - \frac{1}{4}\sigma \tilde{S}^{\gamma \delta} \xi_{\delta;\gamma} u^\alpha u^\beta u^\lambda h^{\mathcal{R}}_{\alpha \beta; \lambda}
    \bigg\rangle.
\end{align}

We wish to manipulate this expression to a tidy form which depends exclusively
on the radiative field, so that we can make a direct comparison with asymptotic flux
amplitudes.
To begin these manipulations, we identify the oscillatory terms that can be related to
covariant derivatives with respect to $\tau$, and remove them via
$\langle D(\dots)/d\tau \rangle = \mathcal{O}(\epsilon^2)$.
Dropping these terms, the orbit-averaged dissipation rate can be simplified to
\begin{align} \label{eq:ExpandedSpinFlux}
  \left\langle \frac{D \gencons}{d \tau} \right\rangle =
  \bigg\langle
  & \frac{1}{2} u^\alpha u^\beta \mathcal{L}_{\xi} h^{\mathcal{R}}_{\alpha \beta}
    + \frac{1}{2} \sigma \xi^\alpha (g_\alpha{}^\lambda +  u_\alpha u^\lambda) u^\beta \tilde{S}^{\delta \epsilon}
       \left(
       h^{\mathcal{R}}_{\lambda\gamma} R^\gamma{}_{\beta \epsilon \delta}
       - h^{\mathcal{R}}_{\beta \delta; \lambda; \epsilon}
       + h^{\mathcal{R}}_{\lambda \delta;\beta; \epsilon}
       + h^{\mathcal{R}}_{\beta \lambda; \delta; \epsilon} \right) \notag\\
  & -\frac{1}{2} \sigma u^\alpha \tilde{S}^{\beta \delta} g^{\gamma \lambda}
    \left( h^{\mathcal{R}}_{\lambda \beta; \alpha} + h^{\mathcal{R}}_{\alpha \lambda; \beta}
    - h^{\mathcal{R}}_{\alpha \beta; \lambda}\right) \xi_{\delta;\gamma}
    \bigg\rangle.
\end{align}
We now take advantage of the symmetries of the Riemann tensor and commute covariant derivatives of the metric perturbation using the standard identity
\begin{equation}
  \nabla_\alpha \nabla_\beta h^{\mathcal{R}}_{\gamma \delta}
  - \nabla_\beta \nabla_\alpha h^{\mathcal{R}}_{\gamma \delta}
  = - h^{\mathcal{R}}_{\gamma \lambda} R^{\lambda}{}_{\delta \alpha \beta}
  - h^{\mathcal{R}}_{\delta \lambda} R^{\lambda}{}_{\gamma \alpha \beta}.
\end{equation}
Via manipulations of the multiple covariant derivatives and symmetries of
Riemann, we re-express the second term of \eqref{eq:ExpandedSpinFlux} as
\begin{equation} \label{eq:SpinFluxSimplifyRiemann}
\tilde{S}^{\delta \epsilon}
  \left(
    h^{\mathcal{R}}_{\lambda\gamma} R^\gamma{}_{\beta \epsilon \delta}
    - h^{\mathcal{R}}_{\beta \delta; \lambda; \epsilon}
    + h^{\mathcal{R}}_{\lambda \delta;\beta; \epsilon}
    + h^{\mathcal{R}}_{\beta \lambda; \delta; \epsilon} \right)
  = \tilde{S}^{\delta \epsilon}
  \left(-h^{\mathcal{R}}_{\gamma \delta} R^\gamma{}_{\epsilon \lambda \beta}
    - h^{\mathcal{R}}_{\beta \delta; \epsilon; \lambda}
    + h^{\mathcal{R}}_{\lambda \delta;\epsilon;\beta} \right).
\end{equation}
Using \eqref{eq:SpinFluxSimplifyRiemann} in
\eqref{eq:ExpandedSpinFlux}, expanding, and integrating by parts for the derivatives
$u^\alpha \nabla_\alpha$, the resulting equation is 
\begin{align}
  \left\langle \frac{D \gencons}{d \tau} \right\rangle =
  \bigg\langle
  & \frac{1}{2} u^\alpha u^\beta \mathcal{L}_{\xi} h^{\mathcal{R}}_{\alpha \beta}
    + \frac{1}{2} \sigma \bigg(-\xi^\lambda u^\beta \tilde{S}^{\delta \epsilon}
    h^{\mathcal{R}}_{\gamma \delta} R^\gamma{}_{\epsilon \lambda \beta}
    - \xi^\alpha u^\beta \tilde{S}^{\delta \epsilon} h^{\mathcal{R}}_{\beta \delta; \epsilon; \alpha}
    - u^\beta \xi^\alpha{}_{;\beta} \tilde{S}^{\delta \epsilon}
    h^{\mathcal{R}}_{\alpha \delta; \epsilon} \bigg) \notag\\
  & -\frac{1}{2} \sigma u^\alpha \tilde{S}^{\beta \delta} g^{\gamma \lambda}
    \left( h^{\mathcal{R}}_{\lambda \beta; \alpha} + h^{\mathcal{R}}_{\alpha \lambda; \beta}
    - h^{\mathcal{R}}_{\alpha \beta; \lambda}\right)  \xi_{\delta;\gamma}
    \bigg\rangle.
\end{align}
Again taking advantage of the Killing vector identity
\eqref{eq:KillingIdent}, we remove another total time derivative using
\begin{equation}
  -  \frac{1}{2} \xi^\lambda u^\beta \tilde{S}^{\delta \epsilon}
  h^{\mathcal{R}}_{\gamma \delta} R^\gamma{}_{\epsilon \lambda \beta}
  - \frac{1}{2 } u^\alpha \tilde{S}^{\beta \delta} g^{\gamma \lambda}
  h^{\mathcal{R}}_{\lambda \beta; \alpha} \nabla_\gamma \xi_\delta
  = -\frac{1}{2} u^\beta \nabla_\beta \left( h^{\mathcal{R}}_{\gamma\delta}
    \tilde{S}^{\delta \epsilon} \nabla^\gamma \xi_\epsilon \right).
\end{equation}
Finally, the remaining terms are equivalent to the covariant derivative of the
Lie derivative of the metric perturbation,
\begin{equation}
  \tilde{S}^{\gamma \delta} u^\beta \nabla_\delta
  \mathcal{L}_{\xi} h^{\mathcal{R}}_{\gamma \beta} =
  \xi^{\alpha } \tilde{S}^{\gamma \delta } u^{\beta }
  h^{\mathcal{R}}_{\beta \gamma;\delta;\alpha}
  -  \tilde{S}_{\gamma }{}^{\delta } u^{\alpha }
  h^{\mathcal{R}}_{\alpha \delta ;\beta} \xi^{\beta }{}^{;\gamma }
  +  \tilde{S}_{\gamma }{}^{\delta } u^{\alpha } \xi^{\beta }{}^{;\gamma }
  h^{\mathcal{R}}_{\alpha \beta;\delta }
  + \tilde{S}^{\gamma \delta } u^{\alpha } \xi^{\beta }{}_{;\alpha }
  h^{\mathcal{R}}_{\beta \gamma }{}_{;\delta } .
\end{equation}
Therefore, when all terms are collected, we conclude that the orbit-averaged
dissipation can be expressed as
\begin{equation} \label{eq:SimpleSpinFlux}
\left\langle \frac{D \gencons}{d\tau} \right\rangle = 
- \frac{1}{2} \left\langle \sigma \tilde{S}^{\gamma \delta} u^\beta \nabla_\delta 
\mathcal{L}_{\xi} h_{\gamma \beta}^{\mathcal{R}} 
- u^\alpha u^\beta \mathcal{L}_{\xi} h_{\alpha \beta}^{\mathcal{R}}  \right\rangle .
\end{equation}
Noting that the time average is the identity operation in the circular orbit case, this reduces in that case to the local piece (i.e. the left-hand side) of the flux-balance law given in Eq.~\eqref{eq:SummaryEquation}.

The equation (\ref{eq:SimpleSpinFlux}) can also be derived by a multipole
expansion of the Dixon-Harte formalism \cite{Dixon-3,Harte:2014wya}.
In particular, the equations presented in Ref.~\cite{Harte:2014wya} give the
expansion in terms of integrals over extended bodies for a more general class of
vectors $\xi$, and for the instantaneous evolution of the quasi-conserved parameters
$\gencons$.
We leave further investigation of the relationship between the powerful
Dixon-Harte formalism for equations of motion and asymptotic fluxes for
future work.

\subsection{Relation to radiative metric perturbations}

While not immediately obvious, it is easy to show that to linear order in $\s$,
Eq.~\eqref{eq:SimpleSpinFlux} in fact depends only on the radiative metric
perturbation.
To see this, we can rewrite the first term in \eqref{eq:SimpleSpinFlux} in terms
of the regular Lorenz-gauge two-point function
\begin{align} \label{eq:SpinFluxGreen1}
  - \frac{1}{2}\left\langle \tilde{S}^{\gamma \delta} u^\beta
  \nabla_\delta \mathcal{L}_{\xi} h_{\gamma \beta}^{\mathcal{R}} \right\rangle 
  &= \frac{1}{2} \frac{\left\langle u^t\right\rangle}{\Delta t} \int_{\Delta t} d^4 x \int_{\Delta t} d^4 x^\prime
    T^{(\s) \alpha \beta} \left(\mathcal{L}_\xi
    G^{\mathcal{R}}_{\alpha \beta \alpha^\prime \beta^\prime}\right)
    T^{\alpha^\prime \beta^\prime} .
\end{align}
Note that we have picked up an orbit-averaged $u^t$ from the ratio of the
implicit period of the time-average operation $\langle\dots\rangle$ (in $\tau$)
and the period $\Delta t$ of the time integration of the Green's function.
We emphasize that this reasoning, like the time-averaging operation itself,
should be treated in the multiscale expansion framework if this procedure is to
be extended to higher order in the mass ratio; our present expansion relies on the
source $T^{\alpha \beta}$ being treated as the instantaneously geodesic source.

The second term in \eqref{eq:SimpleSpinFlux} may be similarly rewritten as
\begin{align} \label{eq:SpinFluxGreen2}
  \frac{1}{2} \left\langle u^\alpha u^\beta
  \mathcal{L}_{\xi} h_{\alpha \beta}^{\mathcal{R}}  \right\rangle 
  &= \frac{1}{2} \frac{\left\langle u^t\right\rangle}{\Delta t} \int_{\Delta t} d^4 x \int_{\Delta t} d^4 x^\prime
    T^{(\m) \alpha \beta} \left(\mathcal{L}_{\xi}
    G^{\mathcal{R}}_{\alpha \beta \alpha^\prime \beta^\prime}\right)
    T^{\alpha^\prime \beta^\prime} .
\end{align}

The defining properties of the regular two-point function, and the fact that $\xi$ is a Killing vector, gives rise to the identity
$\mathcal{L}_\xi G^{\mathcal{R}}_{\alpha \beta \gamma^\prime \delta^\prime}(x,x^\prime) =
-\mathcal{L}_{\xi^\prime} G^{\mathcal{R}}_{\alpha \beta \gamma^\prime
  \delta^\prime}(x,x^\prime)$.
Therefore, the sum of the two expressions (\ref{eq:SpinFluxGreen1}) and
(\ref{eq:SpinFluxGreen2}) depends only on the antisymmetric combination $G^{\mathcal{R}}_{\alpha \beta \alpha^\prime \beta^\prime}(x,x^\prime) - G^{\mathcal{R}}_{\alpha^\prime \beta^\prime \alpha \beta}(x^\prime,x) = G^{\text{Rad}}_{\alpha \beta \alpha^\prime \beta^\prime}(x,x^\prime)$, where $G^{\text{Rad}}_{\alpha \beta \alpha^\prime \beta^\prime}(x,x^\prime)$ is the radiative two-point function.
Thus, to $\mathcal{O}(\s)$, we can rewrite \eqref{eq:SimpleSpinFlux} in terms of
radiative metric perturbations,
\begin{equation} \label{eq:GeoIdentity}
  \left\langle\frac{D \gencons}{d\tau}\right\rangle
  = \frac{1}{2} \left\langle u^\alpha u^\beta \mathcal{L}_\xi \left(h_{\alpha \beta}^{(\mu) \text{Rad}} + 2 \sigma h_{\alpha \beta}^{(\sigma) \text{Rad}}\right) \right\rangle,
\end{equation}
where $h_{\alpha \beta}^{(\mu) \text{Rad}}$ is the radiative part of the perturbation sourced by $T_{\alpha \beta}^{(\mu)}$ and $h_{\alpha \beta}^{(\s) \text{Rad}}$ is the radiative part of the perturbation sourced by $T_{\alpha \beta}^{(\s)}$.

\subsection{Asymptotic fluxes} \label{sec:FluxBalance}

We now complete the derivation of the flux-balance law by relating the sum of terms on
the right-hand sides of Eqs.~\eqref{eq:SpinFluxGreen1} and~\eqref{eq:SpinFluxGreen2}
to the asymptotic mode amplitudes of the metric perturbation.
First, we combine the two terms in a form that emphasizes the symmetries of the equation,
\begin{equation} \label{eq:RadGreensFunctionFlux}
  \left \langle \frac{D \Xi}{d \tau} \right\rangle = \frac{1}{2 \mu} \frac{\left\langle u^t\right\rangle}{\Delta t} \int_{\Delta t} d^4 x \sqrt{-g} \int_{\Delta t} d^4 x^\prime \sqrt{-g} T^{\alpha \beta}(x) \mathcal{L}_\xi G^{\text{Rad}}_{\alpha \beta \alpha^\prime \beta^\prime}(x, x^\prime) T^{ \alpha^\prime \beta^\prime}(x^\prime) + \mathcal{O}(\s^2),
\end{equation}
where the stress energy tensor $T^{\alpha \beta}$ is given by Eq.~\eqref{eq:Texpansion}, and where we have truncated at order $\mathcal{O}(\s^2)$ (note, however, that our expression includes $\mathcal{O}(\s^2)$ contributions necessary to give rise to a nicely symmetric form of the equation).
Note the similarities of this expression to the more general forms derived in
Ref.~\cite{Harte:2014wya}.
Our expression \eqref{eq:RadGreensFunctionFlux} is simpler than Eq.~(216) of
Ref.~\cite{Harte:2014wya} by virtue of our use of a true Killing vector of the
background spacetime, and by our multipole expansion to linear order in the spin
of the small companion.

Equation \eqref{eq:RadGreensFunctionFlux} can be used to relate the rate of
change of orbital quantities $\Xi$ to suitably normalized mode amplitudes of any
set of homogeneous modes for which the radiative two-point function can be written
in a \emph{separated} form,
\begin{equation} \label{eq:GenericSeparableGreenRad}
  G^{\text{Rad}}_{\alpha \beta \alpha^\prime \beta^\prime}(x, x^\prime) = \int d \omega \sum_\Lambda \mathcal{A}_\Lambda i \left[\kappa_{\Lambda \alpha \beta}(x) \bar{\kappa}_{\Lambda \alpha \beta}(x^\prime) e^{i \omega (t - t^\prime)} - \bar \kappa_{\Lambda \alpha \beta}(x) \kappa_{\Lambda \alpha \beta}(x^\prime) e^{i \omega (t^\prime - t)}\right],
\end{equation}
for complex mode functions $\kappa$, $\bar{\kappa}$, collections of mode numbers
$\Lambda$, and normalization constants $\mathcal A_\Lambda$.
The two-point function generated by
(\ref{eq:GenericSeparableGreenRad}) is antisymmetric and real, as required by
the construction of a radiative two-point function for the metric perturbation.

Now, consider the mode decomposition in which the functions
$\kappa_{\Lambda \alpha \beta}$ form a basis for the metric perturbation,
\begin{equation}
  h_{\alpha \beta}(x) = \sum_{\omega} \sum_\Lambda K_{\omega \Lambda} \kappa_{\Lambda \alpha \beta}(x) e^{i \omega t} + \bar{K}_{\omega \Lambda} \bar{\kappa}_{\Lambda \alpha \beta}(x) e^{-i \omega t},
\end{equation}
where we have written the frequency dependence as a sum to emphasize the
discrete spectrum of a bound orbit.
Then, the form of the radiative two-point function gives a simple formula for
the radiative mode amplitudes $K_\Lambda$ of the metric perturbation:
\begin{equation}
  K_{\omega \Lambda} = i \mathcal A_\Lambda \frac{1}{\Delta t} \int_{\Delta t} d^4 x \sqrt{-g} \bar \kappa_{\Lambda \alpha \beta}(x) e^{-i \omega t} T^{\alpha \beta}(x).
\end{equation}
For the final substitution of the two-point function mode expansion into
(\ref{eq:RadGreensFunctionFlux}), we further assume that the modes
$\kappa_{\Lambda \alpha \beta}(x)$ are eigenfunctions of the operator
$\mathcal{L}_\xi$ with eigenvalues $i \lambda_\xi$.
Then, we may re-write the rate of change of the quasiconserved orbital quantities $\Xi$ as,
\begin{equation} \label{eq:GenericModeSum}
  \left \langle \frac{D \Xi}{d \tau} \right\rangle = \sum_\Lambda \frac{\left\langle u^t \right \rangle \lambda_\xi}{\mathcal{A}_\Lambda \mu} |K_\Lambda|^2.
\end{equation}
Note that the $\Delta t$ is cancelled in the final evaluation of the
combination of modes.
 This cancellation can be intuitively understood by considering the symmetric
  expression (\ref{eq:GenericSeparableGreenRad}) as a `total derivative'
  expression on spacetime, giving rise to a flux integrated over the 2+1
  spacetime boundary.
  The cancellation can also be seen more directly by considering the bound orbit
  as possessing the discrete sum of frequency modes as was performed in the
  similar derivation \cite{Sago:2005fn}.

  The above discussion is given to emphasize the generic requirements of the
mode decomposition such that the flux-balance law may be given in terms of a sum
over mode amplitudes.
We now specialize the discussion to the radiation gauge, for which the
separability of the radiative two-point function is well-documented by prior
investigations \cite{Galtsov:1982hwm, Sago:2005fn, Isoyama:2018sib}.
The radiation gauge mode decompositions are defined in terms of the formalism of
metric reconstruction from Teukolsky modes discussed in detail in Sec.~\ref{sec:TeukolskyCalculation}.
The homogeneous modes of the radiation gauge are labeled by their frequency
$\omega$, spin-weighted spheroidal harmonic numbers $\ell$ and $m$, and either
`in' and `up' (corresponding to solutions which vanish at the past horizon 
$\mathcal{H}^-$ and past null infinity $\mathcal{I}^-$, respectively) or `out'
and `down' (which vanish at the future horizon $\mathcal{H}^+$ and future null
infinity $\mathcal{I}^+$, respectively).
In the Teukolsky formalism, the radiative two-point function takes the form,
\begin{align} \label{eq:RadiationGaugeRadGreens}
  G^{\text{Rad}}_{\alpha \beta \alpha^\prime \beta^\prime}(x, x^\prime) &= \text{Re} \int d\omega \frac{1}{i \omega^3} \sum_{\ell m} {}_s \mathcal{A}_{\ell m \omega}^{\text{out}}\, {}_s\pi^{\text{out}}_{\ell m \omega\, \alpha \beta}(x) {}_s \bar \pi^{\text{out}}_{\ell m \omega\,  \alpha^\prime \beta^\prime}(x^\prime) + {}_s \mathcal{A}_{\ell m \omega}^{\text{down}} {}_s \pi_{\ell m \omega\, \alpha \beta}^{\text{down}}(x) {}_s \bar \pi _{\ell m \omega\,  \alpha^\prime \beta^\prime}^{\text{down}}(x^\prime) \notag \\
                                                           &=\text{Re} \int d\omega  \frac{1}{i \omega^3}\sum_{l m}{}_s \mathcal{A}_{\ell m \omega}^{\text{in}} {}_s\pi^{\text{in}}_{\ell m \omega\,  \alpha \beta}(x) {}_s \bar \pi^{\text{in}}_{\ell m \omega\, \alpha^\prime \beta^\prime}(x^\prime) + {}_s \mathcal{A}_{\ell m \omega}^{\text{up}} {}_s \pi_{\ell m \omega\,  \alpha \beta}^{up}(x) {}_s \bar \pi _{\ell m \omega\, \alpha^\prime \beta^\prime}^{\text{up}}(x^\prime).
\end{align}
The mode normalization coefficients ${}_s \mathcal{A}_{\ell m \omega}^{\text{out}}$
and ${}_s \mathcal{A}_{\ell m \omega}^{\text{down}}$ are explicitly derived in
Refs.~\cite{Sago:2005fn, Isoyama:2018sib}, and the corresponding mode normalization
coefficients for in and up modes can be derived by similar methods to those
described in the appendix of Ref.~\cite{Sago:2005fn}.

Defining, then, the mode amplitudes,
\begin{equation}
  {}_s Z_{\ell m \omega}^{\text{in/up}} = i {}_s \mathcal{A}_{\ell m \omega}^{\text{in/up}} \frac{1}{\Delta t} \int_{\Delta t} d^4 x \sqrt{-g} {}_s \bar\pi^{\text{in/up}}_{\ell m \omega\,  \alpha \beta}(x) T^{\alpha \beta}(x),
\end{equation}
and noting that the radiation gauge mode functions are eigenfunctions of
$\mathcal{L}_{\xi_t}$ and $\mathcal{L}_{\xi_\phi}$ with eigenvalues $i \omega$
and $i m$, respectively, we obtain the flux-balance laws,
\begin{subequations}
  \label{eq:fluxbalance}
  \begin{align} \label{eq:energy-fluxbalance}
    \left\langle \frac{D \E}{d \tau}\right\rangle = \frac{\left\langle u^t \right \rangle}{\m}\sum_{\omega} \frac{1}{\omega^2} \sum_{\ell m} \frac{\left|{}_s Z_{\ell m \omega}^{\text{in}}\right|^2}{{}_s \mathcal{A}_{\ell m \omega}^{\text{in}}} + \frac{\left|{}_s Z_{\ell m \omega}^{\text{up}}\right|^2}{{}_s \mathcal{A}_{\ell m \omega}^{\text{up}}} + \mathcal{O}(\s^2)\\
        \left\langle \frac{D L_z}{d \tau}\right\rangle = \frac{\left\langle u^t \right\rangle}{\m}\sum_{\omega} \frac{m}{\omega^3} \sum_{\ell m} \frac{\left|{}_s Z_{\ell m \omega}^{\text{in}}\right|^2}{{}_s \mathcal{A}_{\ell m \omega}^{\text{in}}} + \frac{\left|{}_s Z_{\ell m \omega\, }^{\text{up}}\right|^2}{{}_s \mathcal{A}_{\ell m \omega}^{\text{up}}} + \mathcal{O}(\s^2),
  \end{align}
\end{subequations}
where the amplitudes ${}_s Z_{\ell m \omega}^{\text{in}/\text{up}}$ are understood
to be computed from $T^{ \alpha \beta}$ as given in Eq.~\eqref{eq:Texpansion}. In Sec.~\ref{sec:TeukolskyCalculation} we will explicitly evaluate these mode amplitudes and compute the fluxes for the case of an aligned-spin secondary in a circular orbit about a Schwarzschild black hole.
A similar derivation also follows in the Lorenz gauge, the only caveat being that the equations for
the Lorenz gauge metric perturbation only separate cleanly into modes in the Schwarzschild case. In
that case, one obtains similar expressions for the fluxes in terms of amplitudes of the modes of the
metric perturbation (see, e.g. Sec.~IV of Ref.~\cite{Barack:2005nr} for a derivation of the energy flux). We use this alternative formulation in our Lorenz gauge calculation described in Sec.~\ref{Sec:LorenzGauge}.

\section{Specialization to circular, spin-aligned orbits in Schwarzschild spacetime} \label{Sec:DipoleSource}

Thus far, our discussion has applied to generic orbits in Kerr spacetime. Hereafter, we specialize to the case where the primary is a Schwarzschild black hole of mass $M$, corresponding to a spacetime with line element
\begin{equation}
  ds^2 = -f dt^2 + f^{-1} dr^2 + r^2 (d\theta^2 + \sin^2 d\phi^2),
\end{equation}
where $f \equiv 1-\frac{2M}{r}$. We also specialize to the case where the secondary is moving on
a circular orbit in the equatorial plane, with its spin vector parallel to the orbital angular
momentum. We proceed by first recasting the stress energy into an explicit form which we
further manipulate in Secs.~\ref{sec:Teuk_source} and
\ref{sec:Lorenz-gauge_source} to suit our computational approaches.

\subsection{Circular, spin-aligned orbits in Schwarzschild spacetime}
For a spinning compact object on an aligned, circular, equatorial orbit with radius $r_\p$, the only
non-zero component of the (normalized) spin vector $\tilde{S}^\mu\equiv -\frac12 \epsilon_{\mu \alpha \beta \gamma} u^\alpha \tilde{S}^{\beta \gamma}$ is $\tilde{S}^\theta = -\M/r_\p$. Accordingly, the
(normalized) spin tensor, $\tilde{S}^{\mu\nu} = -\epsilon^{\mu\nu \alpha \beta} \tilde{S}_\alpha u_\beta$ has four non-zero components,
\begin{equation}\label{eq:Smunu_cmpts}
	\tilde{S}^{tr} 		= - \frac{\M}{r_\p} u_\phi =-\tilde{S}^{rt}, \qquad
	\tilde{S}^{r\phi} 	= - \frac{\M}{r_\p} u_t 		= -\tilde{S}^{\phi r}.
\end{equation}
The orbital energy, $\en$, is given by Eq.~\eqref{eq:cons_E} with $\xi = \xi^\mu_t$ and $\Xi = \en$. Writing $\en = \geo{\en} +\s\en_\s$ we have 
\begin{align}
	\geo{\en} = \frac{f_0}{\sqrt{1-3M/r_0}},\qquad \en_\s = -\frac{(M/r_\p)^{5/2}}{\sqrt{1-3M/r_\p}},
\end{align}
where $f_\p = 1-\tfrac{2M}{r_\p}$.
Expanding the orbital frequency through $\mathcal{O}(\s)$, we get
$\O = \Og + \sigma \Os+ \mathcal{O}(\s^2)$, where \cite{Bini:2018zde}
\begin{align}\label{eq:Omega_phi}
	\Og = \sqrt{\frac{\M}{r_\p^3}},\qquad \Os = -\frac{3\M^2}{2 r_\p^3}.
\end{align}
Likewise, expanding the $t$-component of the four-velocity through $\mathcal{O}(\s)$, we obtain $u^t = \geo{u}^t + \s u^t_\s + \mathcal{O}(\s^2)$, where
\begin{align}
	\geo{u}^t = \frac{1}{\sqrt{1-3M/r_\p}},\qquad u^t_\s = -\frac{3M^{5/2}}{2 r_\p (r_\p - 3M)^{3/2}} .
\end{align}

\subsection{Explicit form of the stress-energy}
Starting with Eq.~\eqref{eq:Tmunu} for the stress-energy source, it is convenient to explicitly perform the proper time integration and to expand the dipole term out to yield \cite{Faye:2006gx} 
\begin{subequations}
\label{eq:FBB_spin_source}
\begin{align}
  T^{(\mu)\mu\nu} (t, \mathbf{x}) &= -\frac{1}{\sqrt{-g}}u^\mu V^\nu \delta^3\big[\mathbf{x}-\mathbf{z}(t)\big], \\
	 T^{(\s)\mu\nu} (t, \mathbf{x}) &= -\f{1}{\sqrt{-g}}\left[\pd_\rho\left\{\tilde{S}^{\rho(\mu} V^{\nu)}\delta^3\big[\mathbf{x}-\mathbf{z}(t)\big] \right\} +
	\tilde{S}^{\rho(\mu}\Gamma^{\nu)}_{\rho\sigma}V^\sigma \delta^3\big[\mathbf{x}-\mathbf{z}(t)\big] \right],
\end{align}
\end{subequations}
where $\mathbf{z}(t)=(r_\p, \pi/2, \Omega t)^T$ is the spatial location of the worldline at time $t$ and $V^\alpha\equiv dx^\alpha/dt= u^\alpha/u^t$.
The $\tilde{S}^{\mu \nu}$, $\Gamma^\mu_{\nu \rho}$, $V^\mu$ and $u^\mu$ terms in the spin source are all evaluated on the worldline $\mathbf{z}(t)$, but $g$ is a function of the spacetime coordinates $(t,\mathbf{x})$.

For circular equatorial motion we have $\delta^3\big[\mathbf{x}-\mathbf{z}(t)\big] = \delta_r \delta_\theta \delta_\phi$ , where we have introduced the shorthand $\delta_r \equiv \delta(r-r_\p)$, $\delta_\theta \equiv \delta(\theta-\pi/2)$, $\delta_\phi \equiv \delta(\phi-\Omega t)$. Expanding into components (and noting that we can use $\O = \Og$ in $T^{(\s)\mu\nu}$ since we are working to linear order in $\s$), we get
\begin{subequations}
\label{eq:T_munu}
\begin{align}
T^{(\mu)\mu\nu} &= \f{K_0^{\mu\nu}}{r^2\sin\theta} \delta_r \delta_\theta \delta_\phi,\label{eq:T0_munu_v1} \\
T^{(\s)\mu\nu} &= \f{1}{r^2\sin\theta}\left[K_1^{\mu\nu}\delta_r \delta_\theta \delta_\phi +K_2^{\mu\nu}\delta_r \delta_\theta \delta_\phi'+K_3^{\mu\nu}\delta_r' \delta_\theta \delta_\phi \right],\label{eq:T_munu_v1}
\end{align}
\end{subequations}
where
\begin{equation}
 K_0^{\mu\nu} =  u^\mu V^\nu\Big|_{\mathbf{x}=\mathbf{z}}, \quad
 K_1^{\mu\nu} =- \tilde{S}^{\rho(\mu}\Gamma^{\nu)}_{\rho\sigma}V^\sigma\Big|_{\mathbf{x}=\mathbf{z}},\quad
 K_2^{\mu\nu} =- \big(\tilde{S}^{\phi(\mu}V^{\nu)}-\Og\, \tilde{S}^{t(\mu}V^{\nu)}\big)\Big|_{\mathbf{x}=\mathbf{z}} ,\quad
 K_3^{\mu\nu} =- \tilde{S}^{r(\mu}V^{\nu)}\Big|_{\mathbf{x}=\mathbf{z}}.
\end{equation}
Explicitly, the non-zero components of $K_i^{\mu\nu}$ are
\begin{gather}
	 K_0^{tt} = u^t, \quad K_0^{t\phi} = u^\phi, \quad K_0^{\phi\phi} = (u^\phi)^2/u^t, \nonumber \\
	 K_1^{tt} = \f{-M^{5/2}}{r_\p(r_\p-2\M)\sqrt{r_\p-3\M}},
	 \quad K^{t\phi}_1=K^{\phi t}_1 = -\f{M^2}{r_\p^{5/2}\sqrt{r_\p-3\M}},\nonumber\\
	 K_1^{rr}=-\f{{M}^{3/2}(r_\p-2\M)\sqrt{r_\p-3\M}}{r_\p^3},\quad
	 \quad K_1^{\phi\phi}=-\f{{M}^{3/2}(r_\p -2\M)}{r_\p^4\sqrt{r_\p-3\M}},\nonumber\\
	 K_2^{tr} =K_2^{rt} = \f{\sqrt{r_\p-3\M}}{2r_\p^{3/2}},\quad
	 K_2^{r\phi} = K_2^{\phi r}=\f{\sqrt{M}\sqrt{r_\p-3\M}}{2r_\p^{3}},\nonumber\\
	 K_3^{tt} = -\f{\sqrt{M}}{\sqrt{r_\p-3\M}},\quad
	 K_3^{t\phi} = K_3^{\phi t}= - \f{r_\p-\M}{2r_\p^{3/2}\sqrt{r_\p-3\M}}, \quad
	 K_3^{\phi \phi} = -\f{\sqrt{M}(r_\p-2\M)}{r_\p^{3}\sqrt{r_\p-3\M}}.\label{eq:K_cmpts}
\end{gather}
The total stress-energy is then given by $T^{\mu\nu} = \frac{K^{\mu\nu}}{r^2\sin\theta}$ where
\begin{equation}
  K^{\mu\nu} \equiv \left[(K_0^{\mu\nu}+\s K_1^{\mu\nu})\delta_r \delta_\theta \delta_\phi +\s K_2^{\mu\nu}\delta_r \delta_\theta \delta_\phi'+ \s K_3^{\mu\nu}\delta_r' \delta_\theta \delta_\phi \right].
\end{equation}
Note that the dependence on $(t,\mathbf{x})$ only appears through the prefactor and through
$\delta_r$, $\delta_\theta$ and $\delta_\phi$; the $K_i$ are constants that only depend on $r_\p$
and $M$.

\section{Computation with the Teukolsky formalism and radiation gauge}
\label{sec:TeukolskyCalculation}

In Sec.~\ref{Sec:Results} we will give explicit results demonstrating flux balance using two
largely-independent calculations, one in Lorenz gauge and another using the Teukolsky formalism and
metric reconstruction in radiation gauge. The practical computation of the flux-balance calculation
in the radiation gauge is mostly standard, following the same methodology as in the non-spinning
case. For completeness, we give an overview of the most pertinent points in the procedure below,
and refer the reader to
Refs.~\cite{Hughes:1999bq,Fujita:2004rb,Fujita:2009uz,Keidl:2010pm,vandeMeent:2015lxa,vandeMeent:2017bcc,Kavanagh:2015lva,Kavanagh:2016idg}
for detailed discussions of the practical details both in the post-Newtonian and numerical
contexts, and to
Refs.~\cite{Chrzanowski:1975wv,Kegeles:1979an,Wald:1978vm,Whiting:2005hr,Pound:2013faa} for further
details on the formalism for metric reconstruction. 

\subsection{Specialisation of Teukolsky formalism to Schwarzschild spacetime}

We now specialize the Teukolsky formalism to Schwarzschild spacetime, in which case:
\begin{enumerate}
  \item The Kinnersley tetrad is
    \begin{equation}
      l^\mu=(f^{-1},1,0,0),\quad
      n^\mu=\f{1}{2}(1,-f,0,0),\quad
      m^\mu=\f{1}{\sqrt{2}r}(0,0,1,i\csc\theta),\quad
      \mb^\mu=\f{1}{\sqrt{2}r}(0,0,1,-i\csc\theta),
    \label{eq:Kinnersley_tetrad_Schw}
    \end{equation}
  \item The spin coefficients are
    \begin{equation}
    \rho = -\f{1}{r},\quad \rho'=\f{f}{2r}\,\quad \tau = \tau' = 0, \quad \bpsi_2 = -\frac{M}{r^3}. \label{eq:NP_spin_coefs}
    \end{equation}
  \item The Geroch-Held-Penrose (GHP) \cite{Geroch:1973am} derivative operators are
    \begin{alignat}{4}
      \th &= f^{-1} \partial_t + \partial_r, \quad
      &\th' &= \tfrac12 (\partial_t -f \partial_r-2 b M/r^2),\nonumber \\
      \edth &= \tfrac{1}{\sqrt{2}r}( \partial_\theta + i \csc \theta \partial_\phi - s \cot\theta),\quad
      &\edth' &= \tfrac{1}{\sqrt{2}r}( \partial_\theta - i \csc \theta \partial_\phi + s \cot\theta),
    \end{alignat}
    where $s$ and $b$ are, respectively, the spin-weight and boost-weight of the quantity being acted on.
  \item The Teukolsky equations for the Weyl scalars (i.e. the tetrad projections of the Weyl tensor, $\ppsi_0 \equiv C_{lmlm}$ and $\ppsi_4 \equiv C_{n\mb n\mb}$) are separable using the ansatz
\begin{align}
  \ppsi_0 &= \int_{-\infty}^\infty d\omega \sum_{\ell=2}^\infty \sum_{m=-\ell}^\ell \, {}_2 \ppsi_{\ell m \omega}(r) \, {}_2 Y_{\ell m}(\theta, \phi) e^{-i \omega t}, \\
  \bpsi_2^{-4/3} \ppsi_4 &= \int_{-\infty}^\infty d\omega \sum_{\ell=2}^\infty \sum_{m=-\ell}^\ell  \,{}_{-2} \ppsi_{\ell m \omega}(r) \, {}_{-2} Y_{\ell m}(\theta, \phi) e^{-i \omega t}.
\end{align}
  \item The spin-weighted spherical harmonics, ${}_s Y_{\ell m}(\theta, \phi)$, satisfy the equation
  \begin{equation}
    \label{eq:SWSH}
      \bigg[\dfrac{d}{d\chi} \bigg((1-\chi^2)\dfrac{d}{d\chi} \bigg)
      -\frac{(m+s \chi)^2}{1-\chi^2} +s + {}_s \lambda_{\ell m} \bigg] {}_{s} Y_{\ell m} = 0,
    \end{equation}
    where $\chi \equiv \cos \theta$, and where the eigenvalue is ${}_s \lambda_{\ell m} = \ell(\ell+1) - s(s+1)$. They are unit-normalised on the sphere, $\int {}_s Y_{\ell m} (\theta, \varphi) {}_s \bar{Y}_{\ell' m'}(\theta, \varphi) {\rm d} \Omega = \delta_{\ell \ell'} \delta_{m m'}$.
  \item The radial functions ${}_s R_{\ell m \omega}$ satisfy the Teukolsky radial equation,
    \begin{equation}
      \label{eq:TeukolskyR}
      \bigg[\Delta^{-s} \dfrac{d}{dr}\bigg( \Delta^{s+1}\dfrac{d }{dr}\bigg)
    +\frac{K^2 - 2 i s (r-M)K}{\Delta} + 4 i s \omega r - {}_s \lambda_{\ell m} \bigg]{}_{s} \ppsi_{\ell m \omega} = {}_{s} T_{\ell m \omega},
    \end{equation}
    where $\Delta \equiv r(r-2M)$ and $K\equiv r^2 \omega$.
  \item We work with a basis of radiative homogeneous solutions, ${}_s R^{\text{in}}_{\ell m \omega}$ and ${}_s R^{\text{up}}_{\ell m \omega}$, which vanish at $\mathcal{H}^-$ and $\mathcal{I}^-$, respectively. We choose to normalise these such that transmission coefficients are $1$. Our homogeneous therefore have the asymptotic behaviour
\begin{subequations}
\begin{alignat}{3}
\label{eq:bcRin}
{}_s R^{\text{in}}_{\ell m \omega}(r) &\sim
\left\{
\begin{array}{rcll}
 0&+&
 {}_s R^{\text{in,trans}}_{\ell m \omega} \Delta^{-s} e^{-i \omega r_*}
 \\ \quad
 {}_s R^{\text{in,ref}}_{\ell m \omega} r^{-1-2s} e^{+i\omega r_*}
 &+&
 {}_s R^{\text{in,inc}}_{\ell m \omega} r^{-1} e^{-i\omega r_*}
\end{array}
\right.
&\qquad
\begin{array}{l}
  r \to r_+\\
  r \to \infty
\end{array},
\\
\label{eq:bcRup}
{}_s R^{\text{up}}_{\ell m \omega}(r) &\sim
\left\{
\begin{array}{rcll}
 {}_s R^{\text{up,inc}}_{\ell m \omega} e^{+i \omega r_*}
 &+&
 {}_s R^{\text{up,ref}}_{\ell m \omega} \Delta^{-s} e^{-i \omega r_*}
 \\
 {}_s R^{\text{up,trans}}_{\ell m \omega} r^{-1-2s} e^{+i\omega r_*}
 &+&
 0
\end{array}
\right.
&\qquad
\begin{array}{l}
  r \to r_+\\
  r \to \infty
\end{array} ,
\end{alignat}
\end{subequations}
where $r_* = r + 2 M \ln \frac{r-r_\mathcal{H}}{2M}$ and $r_\mathcal{H} \equiv 2M$.
  \item When acting on the spin-weighted spherical harmonics, $\edth$ and $\edth'$ are essentially
    spin-raising and lowering operators,
    \begin{subequations}
    \begin{align}
      \sqrt{2} r \,\edth \big[{}_s Y_{\ell m}(\theta, \phi)\big] &= -\big[\ell (\ell + 1) - s (s + 1)\big]^{1/2} {}_{s+1} Y_{\ell m} (\theta, \phi),\\
      \sqrt{2} r \, \edth' \big[{}_s Y_{\ell m}(\theta, \phi)\big] &= \big[\ell (\ell + 1) - s (s - 1)\big]^{1/2} {}_{s-1} Y_{\ell m} (\theta, \phi).
      \end{align}
    \end{subequations}
      Complex conjugating, we have the related identities
      \begin{subequations}
      \begin{align}
        \sqrt{2} r \,\edth \big[{}_s \bar{Y}_{\ell m}(\theta, \phi)\big] &= \big[\ell (\ell + 1) - s (s - 1)\big]^{1/2} {}_{s-1} \bar{Y}_{\ell m} (\theta, \phi),\\
        \sqrt{2} r \, \edth' \big[{}_s \bar{Y}_{\ell m}(\theta, \phi)\big] &= -\big[\ell (\ell + 1) - s (s + 1)\big]^{1/2} {}_{s+1} \bar{Y}_{\ell m} (\theta, \phi).
      \end{align}
      \end{subequations}
  \item The Teukolsky-Starobinsky identities (valid in regions where $\ppsi_0$ and $\ppsi_4$ satisfy the homogeneous Teukolsky equation) yield identities relating the positive spin-weight spheroidal and radial functions to the negative spin-weight ones \cite{Chandrasekhar:1985kt,Ori:2002uv},
\begin{subequations}
  \label{eq:TS-mode}
\begin{align}
 \mathcal{D}_0^4 ({}_{-2} \ppsi_{\ell m \omega}) &= \tfrac14 \mathcal{C}_{\ell m \omega}  \, {}_{2} \ppsi_{\ell m \omega}, \\
 \Delta^{2} (\mathcal{D}^\dag_0)^4 (\Delta^{2}\, {}_{2} \ppsi_{\ell m \omega}) &= 4 \bar{\mathcal{C}}_{\ell m \omega}  \,{}_{-2} \ppsi_{\ell m \omega}, \\
 \mathcal{L}_{-1}\mathcal{L}_{0}\mathcal{L}_{1}\mathcal{L}_{2} ({}_{2} Y_{\ell m}) &= D \, {}_{-2} Y_{\ell m}, \\
 \mathcal{L}^\dag_{-1}\mathcal{L}^\dag_{0}\mathcal{L}^\dag_{1}\mathcal{L}^\dag_{2} ( {}_{-2} Y_{\ell m}) &= D \, {}_{2} Y_{\ell m},
\end{align}
\end{subequations}
where $\mathcal{D}_n \equiv \partial_r - \frac{i K}{\Delta} + 2 n \frac{r-M}{\Delta}$,
$\mathcal{D}^\dag_n \equiv \partial_r + \frac{i K}{\Delta} + 2 n \frac{r-M}{\Delta}$,
$\mathcal{L}_n \equiv \partial_\theta + m
\csc \theta + n \cot \theta$ and $\mathcal{L}^\dag_n \equiv
\partial_\theta - m
\csc \theta + n \cot \theta$ are essentially mode versions of the GHP differential operators.
The constants of proportionality are given by $\mathcal{C}_{\ell m \omega} = D + (-1)^{\ell+m} 12
i M \omega$ and $D=(\ell-1)\ell(\ell+1)(\ell+2)$.
This particular choice of $\mathcal{C}_{\ell m \omega}$ ensures that the $s=+2$ and $s=-2$ modes represent the same physical perturbation.\footnote{An 
alternative proportionality constant can be derived such that the $s=+2$ and $s=-2$ modes have the same
transmission coefficient; see \cite{Ori:2002uv} for details.}
\item Inhomogeneous solutions of the radial Teukolsky equation can constructed from a linear combination of the basis functions,
\begin{align} \label{eq:TeukolskyInhomogeneousModes}
  {}_2 \ppsi_{\ell m \omega}(r) &= {}_2
    C^{\text{in}}_{\omega \ell m}(r) {}_2 R^{\text{in}}_{\ell m \omega}(r)+{}_2
    C^{\text{up}}_{\omega \ell m}(r) {}_2 R^{\text{up}}_{\ell m \omega}(r), \\
  {}_{-2} \ppsi_{\ell m \omega}(r) &= {}_{-2}
    C^{\text{in}}_{\omega \ell m}(r) {}_{-2} R^{\text{in}}_{\ell m \omega}(r)+{}_{-2}
    C^{\text{up}}_{\omega \ell m}(r) {}_{-2} R^{\text{up}}_{\ell m \omega}(r),
\end{align}
where the weighting coefficients are determined by the variation of parameters,
\begin{subequations}
  \label{eq:weighting-coefficients}
  \begin{align}
  {}_s C_{\omega \ell m}^{\text{in}}(r) &= \int^{r_\mathcal{I}}_r \frac{{}_s R^{\text{up}}_{\ell m \omega}(r')}{W(r')\Delta} {}_s T_{\ell m \omega}(r') dr', \\
  {}_s C_{\omega \ell m}^{\text{up}}(r) &= \int_{r_\mathcal{H}}^r \frac{{}_s R^{\text{in}}_{\ell m \omega}(r')}{W(r')\Delta} {}_s T_{\ell m \omega}(r') dr',
\end{align}
\end{subequations}
where $W(r) = {}_s R^{\text{in}}_{\ell m \omega}(r) \partial_r [{}_s R^{\text{up}}_{\ell m \omega}(r)] - {}_s R^{\text{up}}_{\ell m \omega}(r) \partial_r [{}_s R^{\text{in}}_{\ell m \omega}(r)]$ is the Wronskian [in practice, it is convenient to use the fact that $\Delta^{s+1} W(r) = \text{const}$].
\item The fluxes of energy through infinity and the horizon can be determined from the ``in'' and ``up'' normalization coefficients \cite{Hughes:1999bq},
\begin{subequations}
\label{eq:flux-modes}
\begin{align}
  \fluxH &= 2\sum_{\ell=2}^\infty \sum_{m=1}^\ell \alpha_{\ell m \omega} \frac{|2\pi\, {}_{-2} C^{\text{in}}_{\ell m \omega}(r_\mathcal{H})|^2}{4\pi \omega^2}, \\
  \fluxI &= 2\sum_{\ell=2}^\infty \sum_{m=1}^\ell \frac{|2\pi\, {}_{-2} C^{\text{up}}_{\ell m \omega}(r_\mathcal{I})|^2}{4\pi \omega^2},
\end{align}
\end{subequations}
where $\alpha_{\ell m \omega} \equiv \frac{256(2M r_\mathcal{H})^5 (\omega^2+4\varepsilon^2)(\omega^2+16\varepsilon^2)\omega^4}{|\mathcal{C}_{\ell m \omega}|^2}$ with 
$\varepsilon \equiv \frac{1}{4 r_\mathcal{H}}$.
\item 
Solutions of the Teukolsky equation can be related
back to solutions for the metric perturbation $h_{\alpha \beta}$ by use of a Hertz potential
 \cite{Wald:1978vm, Chrzanowski:1975wv, Kegeles:1979an, Lousto:2002em,
Whiting:2005hr}. In fact, there are two different Hertz potentials: $\hpsi^{\rm IRG}$, which produces a metric perturbation in the ingoing radiation gauge (satisfying $l^\alpha h_{\alpha\beta}=0$ and $h=0$); and $\hpsi^{\rm ORG}$, which produces a metric perturbation in the outgoing radiation gauge (satisfying $n^\alpha h_{\alpha\beta}=0$ and $h=0$).

In the outgoing radiation gauge (ORG), the metric perturbation may be written in terms of a second-order differential operator acting on a GHP type $\{4,0\}$ (i.e. $s=b=2$, the same as $\ppsi_0$) Hertz potential, $\hpsi^{\rm IRG}$.\footnote{Some authors \cite{vandeMeent:2015lxa} define a slightly different ORG Hertz
potential related to ours by $\hat{\hpsi}^{\rm ORG} = \bpsi_2^{4/3} \hpsi^{\rm ORG}$ and
$(\hat{\mathcal{S}}_4^{\mu\nu})^\dag = (\mathcal{S}_4^{\mu\nu})^\dag \bpsi_2^{-4/3}$. Both
conventions yield the same metric perturbation, $(\hat{\mathcal{S}}_4^{\mu\nu})^\dag
\hat{\hpsi}^{\rm ORG} = (\mathcal{S}_4^{\mu\nu})^\dag \hpsi^{\rm ORG}$.} In terms of this Hertz potential, the ORG metric perturbation is given explicitly by $h_{\mu\nu}^{\rm ORG} =
\Re[ (\mathcal{S}^{\alpha \beta})^\dag \hpsi^{\rm ORG}]$, where
\begin{align}
  (\mathcal{S}_4^{\alpha \beta})^\dag &= n^\alpha n^\beta (\edth'-\tau')(\edth'+3\tau') +\mb^{\alpha}\mb^\beta (\th'-\rho')(\th' + 3\rho') \nonumber \\
  &  \qquad- n^{(\alpha}\bar{m}^{\beta)} \big[(\th'-\rho'+\bar{\rho}')(\edth'+3\tau') +(\edth'-\tau'+\bar{\tau})(\th'+3\rho')]
\end{align}
is the adjoint of $\mathcal{S}_4^{\alpha\beta}$ (given below) and where $\bpsi_2^{4/3}\hpsi^{\rm ORG}$ is a solution of the equation satisfied by $\ppsi_0$ (equivalently, the adjoint of the equation satisfied by $\bpsi_2^{-4/3}\ppsi_4$), but with a different source. 

The ORG Hertz potential may be obtained either by solving this sourced Teukolsky equation or by solving either one of a pair of fourth-order differential equations sourced by the perturbed Weyl scalars, often referred to as the ``angular'' and ``radial'' inversion equations. In regions where $\ppsi_0$ satisfies the homogeneous Teukolsky equation, the ORG Hertz potential satisfies a homogenous equation and the angular inversion equation simplifies significantly, to the point where it can be inverted algebraically. When written in terms of modes, this gives the modes of the ORG Hertz potential, $\hpsi^{\text{ORG}}_{\ell m \omega}$, in terms of the modes of the Weyl scalar,
\begin{align} \label{eq:angular-inversion-ORG}
&\hpsi^{\text{ORG}}_{\ell m\omega}= 
 8\frac{(-1)^m D\;\! {}_2\bar{\ppsi}_{-\omega\ell-m}+12 i M \omega\;\! {}_2\ppsi_{\omega\ell m} }{|\mathcal{C}_{\ell m \omega}|^2}.
\end{align}

\end{enumerate}

\subsection{Explicit source for the Teukolsky equation}
\label{sec:Teuk_source}

We now construct the explicit expressions for the source for the Teukolsky equation for a spinning
secondary in a circular orbit around a Schwarzschild black hole with its spin parallel to the
orbital angular momentum. To do so, we apply the operator\footnote{In fact, there are two operators $\mathcal{S}^{\alpha\beta}_0$ and $\mathcal{S}^{\alpha\beta}_4$ which produce sources for $\ppsi_0$ (of spin-weight $s=+2$) and $\ppsi_4$ (of spin-weight $s=-2$), respectively.} $\mathcal{S}^{\alpha\beta}$ given by
\begin{subequations}
\label{eq:S}
\begin{align}
\mathcal{S}^{\alpha\beta}_0 &= 
  (\edth-\bar{\tau}'-4\tau) 
    \big[(\th-2\bar{\rho})l^{(\alpha} m^{\beta)}-(\edth-\bar{\tau}') l^\alpha l^\beta \big]
  + (\th-4\rho-\bar{\rho})
    \big[(\edth-2\bar{\tau}') l^{(\alpha} m^{\beta)} - (\th-\bar{\rho}) m^\alpha m^\beta \big], \\
\mathcal{S}^{\alpha\beta}_4 &= 
  (\edth'-\bar{\tau}-4\tau') 
    \big[(\th'-2\bar{\rho}')n^{(\alpha} \mb^{\beta)}-(\edth'-\bar{\tau}) n^\alpha n^\beta \big]
  + (\th'-4\rho'-\bar{\rho}')
    \big[(\edth'-2\bar{\tau}) n^{(\alpha} \mb^{\beta)} - (\th'-\bar{\rho}') \mb^\alpha \mb^\beta \big]
\end{align}
\end{subequations}
to the stress-energy tensor given in Eq.~\eqref{eq:T_munu} then decompose into spin-weighted spherical harmonic and Fourier modes,
\begin{equation}
  {}_s T_{\ell m \omega} = - 4 \int_{-\infty}^\infty e^{i \omega t} \int_0^\pi \int_0^{2\pi} \,{}_s \bar{S}_{\ell m}(\theta, \phi) \bpsi_2^{(s-2)/3} \mathcal{S}^{\alpha \beta}_s T_{\alpha \beta} \, \Sigma \sin\theta\, d\theta\, d\phi\, dt.
\end{equation}
In doing so, we exploit the fact that angular derivatives (which appear via $\edth$ and $\edth'$ in $\mathcal{S}^{\alpha \beta}_s$) can be shifted onto the harmonic by integrating by parts. This is particularly simple in the Schwarzschild case, where $\tau = 0 = \tau'$ so that the adjoints of the operators are given by $\edth^\dag = - \edth$ and $(\edth')^\dag=-\edth'$. We therefore have the identities
\begin{subequations}
\begin{align}
  \sqrt{2}\, r \int {}_s \bar{Y}_{\ell m}(\theta, \phi) \mathcal{\edth} X(\theta,\phi) \,\sin\theta\, d\theta\, d\phi = -\big[\ell (\ell + 1) - s (s - 1)\big]^{1/2} \int {}_{s-1} \bar{Y}_{\ell m}(\theta, \phi)X(\theta,\phi) \,\sin\theta\, d\theta\, d\phi \label{eq:int_by_parts},\\
  \sqrt{2}\, r \int {}_s \bar{Y}_{\ell m}(\theta, \phi) \mathcal{\edth}' X(\theta,\phi) \,\sin\theta\, d\theta\, d\phi = \big[\ell (\ell + 1) - s (s + 1)\big]^{1/2} \int {}_{s+1} \bar{Y}_{\ell m}(\theta, \phi)X(\theta,\phi) \,\sin\theta\, d\theta\, d\phi\label{eq:int_by_parts_prime}
\end{align}
\end{subequations}
for any sufficiently smooth function $X(\theta,\phi)$ of appropriate type such that the integrand has zero spin-weight.

Using the fact that the projection of the stress-energy onto the Kinnersley tetrad is given by
\begin{gather}
 T_{ll} = \frac{\mu}{r^2 f^2 \sin\theta}\Big[f^2 K^{tt} -2 f K^{tr} + K^{rr}\Big],\quad
 T_{nn} = \frac{\mu}{4r^2\sin\theta}\Big[f^2 K^{tt} +2 f K^{tr} + K^{rr}\Big],\nonumber \\
 T_{lm} = -\f{i \mu}{\sqrt{2}\, r f} \Big[ f K^{t \phi} - K^{r \phi}\Big],  \quad
 T_{n\bar{m}} = \f{i\mu }{2\sqrt{2}\, r} \Big[ f K^{t \phi} + K^{r \phi}\Big], \quad
 T_{\bar{m}\bar{m}} =T_{mm}=-\f{\mu \sin\theta}{2} K^{\phi\phi}, \label{eq:T-proj}
\end{gather}
we obtain an expression for the source for the Teukolsky equation of the form
\be
{}_s T_{\ell m \omega} =\mu \left[{}_s T^{(0)}_{\ell m \omega}+\ {}_s T^{(1)}_{\ell m \omega}+\ {}_s T^{(2)}_{\ell m \omega} \right],\label{eq:splus2_source_FD}
\ee
along with the condition $\omega = m\Omega$ which follows from the $t$-integral. The individual terms in the $s=+2$ case are given by
\begin{subequations}
\label{eq:splus2_T}
\begin{align}
{ }_2 T^{(0)}_{\ell m \omega} &= \f{2}{f^2 r^2}\sqrt{(\ell-1)\ell(\ell+1)(\ell+2)}\, {}_0\bar{Y}_{\ell m}(\tfrac{\pi}{2},0)\Big[ \big(f^2 K_{01}^{tt}+\s K_1^{rr}-2 i m \s f K_2^{tr}\big)\delta_r+\s f^2K_3^{tt}\delta'_r\Big],\label{eq:splus2_T0}\\
{ }_2 T^{(1)}_{\ell m \omega} &= -8 i  \sqrt{(\ell-1)(\ell+2)}\, {}_1\bar{Y}_{\ell m}(\tfrac{\pi}{2},0) \Big[
\big(F_3 K_{01}^{t\phi}-\s F_4 K_2^{r\phi}\big) \delta_r 
+\big(\tfrac{1}{2} K_{01}^{t\phi}-\tfrac{i m\s}{2f}  K_2^{r\phi}+ \s F_3 K_3^{t\phi} \big) \delta'_r + \tfrac{\s}{2}  K_3^{t\phi} \delta''_r \Big] ,\label{eq:splus2_T1}\\
{ }_2 T^{(2)}_{\ell m \omega} &=- 2 r^2 \,{}_2\bar{Y}_{\ell m}(\tfrac{\pi}{2},0) \Big[ 
F_1 K_{01}^{\phi\phi} \delta_r +\big(F_2 K_{01}^{\phi\phi} + \s F_1 K_3^{\phi\phi} \big) \delta'_r
+\big(K_{01}^{\phi\phi} + \s F_2 K_3^{\phi\phi} \big) \delta''_r+ \s K_3^{\phi\phi} \delta'''(r-r_\p)
\Big] \label{eq:splus2_T2},
\end{align}
\end{subequations}
where we have introduced the shorthand $K_{01}^{\mu\nu} \equiv K_0^{\mu\nu}+\s K_1^{\mu\nu}$, and where
\begin{equation}
 F_1 \equiv \f{4}{r^2} + i \omega \left(\f{f'}{f^2}-\f{6}{r f}\right)-\f{\omega^2}{f^2}, \quad
 F_2 \equiv 2\left(\f{3}{r}-\f{i \omega}{f}\right),\quad
 F_3 \equiv \f{1}{r}-\f{i \omega}{2f},\quad
 F_4 \equiv \f{im}{f^2}\left(\f{r-3M}{r^2}-\f{i \omega}{2} \right),
\end{equation}
with $f' = \partial_r f = 2\M/r^2$. Note that whereas $K^{\mu\nu}_{1,2,3}$ are constant in $r$, the $F_i$ are functions of $r$. Similarly, the terms for $s=-2$ are
\begin{subequations}
\begin{align}
{ }_{-2}T^{(0)}_{\ell m \omega} &= \frac{ r^2}{2}\sqrt{(\ell-1)\ell(\ell+1)(\ell+2)}\, {}_0 \bar{Y}_{\ell m}(\tfrac{\pi}{2},0)\Big[ \big(f^2 K_{01}^{tt}+\s K_1^{rr}+2im \s f K^{tr}_2\big)\delta_r+\s f^2{K}_3^{tt}\delta'_r\Big],\label{eq:sminus2_T0}\\
{ }_{-2}T^{(1)}_{\ell m \omega} &=  2 i f^2 r^4 \sqrt{(\ell-1)(\ell+2)} \, {}_{-1} \bar{Y}_{\ell m}(\tfrac{\pi}{2},0) \Big[
(\bar{F_3} K_{01}^{t\phi}-\s \bar{F}_4 K_2^{r\phi}) \delta_r
+\big(\tfrac{1}{2} K_{01}^{t\phi}+\tfrac{i m\s}{2f}  K_2^{r\phi}+ \s \bar{F}_3 K_3^{t\phi} \big) \delta'_r+\tfrac{\s}{2}  K_3^{t\phi}\delta''_r\Big],\label{eq:sminus2_T1}\\
{ }_{-2}T^{(2)}_{\ell m \omega} &=-\frac{ f^2 r^6 }{2}\, {}_{-2} \bar{Y}_{\ell m}(\tfrac{\pi}{2},0) \Big[
 \bar{F}_1 K_{01}^{\phi\phi}\delta_r +\big(\bar{F}_2 K_{01}^{\phi\phi}+ \s \bar{F}_1 K_3^{\phi\phi} \big) \delta'_r
+\big(K_{01}^{\phi\phi}+ \s \bar{F}_2 K_3^{\phi\phi}\big) \delta''_r+ \s K_3^{\phi\phi} \delta'''_r
\Big] \label{eq:sminus2_T2}.
\end{align}
\end{subequations}

Inserting these into the variation of parameters integral, we obtain expressions for the weighting coefficients, ${}_s C^{\text{in}}_{\ell m \omega}(r) = \theta(r_\p-r){}_s C^{\text{in}}_{\ell m \omega}(r_\p)$ and ${}_s C^{\text{up}}_{\ell m \omega}(r) = \theta(r-r_\p){}_s C^{\text{up}}_{\ell m \omega}(r_\p)$ where the coefficients for both ``in'' and ``up'' can be found in Appendix \ref{apdx:weighting_coeffs}.

\subsection{Numerical and analytical solutions to the Teukolsky equation}

\subsubsection{Post-Newtonian calculation in the small mass-ratio limit}

Our post-Newtonian solutions to the Teukolsky equation are formed by making the following assumptions
\begin{enumerate}[label=(\roman*)]
	\item 
	$r\sim r_\p\gg M$ \\
	Physically, this implies that the small body is at all times far from the central black hole, and that we are calculating the field near this radius. For example our solutions will not be valid in the regime $r\gg r_\p\gg M$.
	\item
	 $\omega\propto\O\sim r_\p^{-3/2}$ \\
	 This is required when the small body moves on a bound (in our case circular) orbit. 
	 As one would expect for a periodic orbit, this condition is implicitly enforced by setting the allowed frequencies to be multiples of the orbital frequency $\omega=m \O$, where $\O$ is given by Eq.~\eqref{eq:Omega_phi}.
\end{enumerate}
These two assumptions are sufficient to construct analytic solutions to the Teukolsky equation as an asymptotic expansion in $u=\frac{M}{r_\p}$ following procedures outlined in, e.g., Refs.~\cite{Kavanagh:2015lva,Kavanagh:2016idg}.
The main difference with many previous works is that at each order in the expansion, we will also introduce an expansion in the dimensionless spin of the small black hole $\s$, which enters via the frequencies of the homogeneous solutions, and with the source terms when constructing inhomogeneous solutions. 
For reasons outlined above we work to linear order in $\s$. 
As is standard, we will exchange the `gauge dependent' expansion parameter $u$ with the more physical frequency variable $y=(M\O)^{2/3}$. 
This is a seemingly arbitrary choice in this work, but would play a more significant role when working either to higher orders in the mass-ratio, with quantities which are not zero as the mass-ratio goes to zero, with gauge-dependent quantities, or when working with the effective-one-body approach, see for example Refs.~\cite{Nagar:2019wrt,Bini:2018zde}.

\subsubsection{Numerical computation}

The numerical computation follows a very similar path as the post-Newtonian calculation outlined above, with the main exception that the homogeneous solutions to the Teukolsky equation are computed numerically. 
In practice we compute the $s=-2$ homogeneous solutions using the semi-analytic MST method (see Ref.~\cite{Sasaki:2003xr} for a review of the formalism and Refs.~\cite{Fujita:2004rb, Fujita:2009uz} for discussion of numerical techniques we employ). 
Once the homogeneous solutions are in hand, the inhomogeneous solutions are constructed via the standard variation of parameters approach -- see Eqs.~\eqref{eq:weighting-coefficients}. 
From the inhomogeneous solutions, the asymptotic energy flux at the spacetime boundaries can be computed using Eq.~\eqref{eq:flux-modes}. 
To reconstruct the metric, we first transform the $s=-2$ Teukolsky solutions to the $s=+2$ solutions using the Teukolsky-Starobinsky identities, Eq.~\eqref{eq:TS-mode}\footnote{This is a legacy step required to connect two pieces of code, one that computes $s=-2$ Teukolsky solutions and another that computes the metric perturbation from the $s=+2$ solutions.}. 
The metric is then reconstructed in the ORG as described above and the change to the local energy is computed via Eq.~\eqref{eq:SimpleSpinFlux}. 
The numerical Teukolsky code uses arbitrary precision throughout as this is required by some pieces of the MST calculation.

\section{Computation in the Lorenz gauge} \label{Sec:LorenzGauge}

The Lorenz gauge has been heavily employed in self-force computations as the original regularization procedure was formulated in this gauge \cite{Mino:1996nk,Quinn:1996am}.
It also has the advantage of working directly with the metric perturbation, thus avoiding the complicated metric reconstruction procedure required when working with the Teukolsky formalism.
This comes at the expense of having to solve a coupled set of ODEs unlike in the Teukolsky case, where one solves for a single variable master function.

Lorenz gauge computations have been carried out in the time domain with 1+1 \cite{Barack:2002ku, Barack:2005nr, Barack:2007tm, Barack:2010tm} and 2+1 \cite{Dolan:2012jg, Isoyama:2014mja} dimensional decompositions as well as in the frequency domain \cite{Akcay:2010dx, Akcay:2013wfa, Osburn:2014hoa, Wardell:2015ada}.
In this work, we employ the frequency-domain approach.
For a spinning body, the decomposition of the metric perturbation into tensor spherical and frequency modes is the same as for the geodesic case, and as such we only briefly review this below. 
The mode decomposition of the source for a spinning body, however, is new and we discuss this in detail before a brief review of our numerical scheme.

In our setup, an orbiting particle of mass $\m$ and spin $\s$ induces a metric perturbation $h_{\mu\nu}$ over the background (Schwarzschild) spacetime, $g_{\mu\nu}$.
We will find it convenient to write the associated field equations with respect to the trace-reversed metric perturbation, $\bar{h}_{\mu\nu}$, given by
\begin{align}
	\hb_{\mu\nu} \equiv h_{\mu\nu} - \frac{1}{2}g_{\mu\nu} \text{Tr}(h).
\end{align}
With this the Lorenz-gauge condition is given by
\begin{align}\label{eq:Lorenz-gauge_condition}
	\nabla_\mu \hb^{\mu\nu} = 0,
\end{align}
where $\nabla$ is the covariant derivative with respect to the background metric.
Applying the gauge condition to the field equations, we get the Lorenz-gauge linearized Einstein equation,
\begin{align}\label{eq:linearized_Einstein_Lorenz_gauge}
	\square \hb_{\mu\nu} + 2 \tensor{R}{^\rho_\mu^\sigma_\nu}\,\hb_{\rho\sigma} = -16\pi T_{\mu\nu},
\end{align}
where $\square = \nabla_\mu \nabla^\mu$, $R$ is the Riemann tensor of the background spacetime, and $T$ is the stress-energy tensor given in Eq.~\eqref{eq:FBB_spin_source}.

We proceed by decomposing $\hb_{\mu\nu}$ onto a basis of tensor spherical harmonics and Fourier modes.
For circular orbits the Fourier mode frequencies are discrete, being given by $\omega \equiv \omega_m = m\Omega$. 
Thus the integral over $\omega$ in the standard Fourier decomposition reduces to a sum over $m$ in this case.
Therefore we may expand $\hb_{\mu\nu}$ as
\begin{align}\label{eq:hb_decomp}
	\hb_{\mu\nu} = \frac{\mu}{r} \sum_{\l,m}\sum_{i=1}^{10} a^{(i)}_\l  \hb_{\l m}^{(i)}(r) Y_{\mu\nu}^{(i)\l m}(\theta,\varphi; r) e^{-i\omega_m t},
\end{align}
where $Y_{\mu\nu}^{(i)\l m}$ form a tensor spherical harmonic basis with $i=1\dots10$ and the $a_\l^{(i)}$ are $\l$-dependent factors, with both given explicitly in Appendix A of Ref.~\cite{Wardell:2015ada}.
The decomposition of the source is similar and is discussed in detail in the next section.
Substituting the decomposition \eqref{eq:hb_decomp} into the field equations \eqref{eq:linearized_Einstein_Lorenz_gauge} results in separable equations. The spherical symmetry of the background geometry ensures that the $\l m$-modes decouple, though in general within each $\l m$-mode a subset of the $i$-fields remain coupled. The resulting radial equation takes the form
\begin{align}\label{eq:Lorenz_gauge_FD_ODE}
	\square^\text{sc}_{\l m} \hb^{(i)}_{\l m} - 4f^{-2}\tensor{\mathcal{M}}{^{(i)}_{\!\!(j)}} \hb^{(j)}_{\l m} = \mathcal{J}^{(i)}_{\l m},
\end{align}
where $\square^\text{sc}_{\l m}$ is the scalar wave operator
\begin{align}
	\square^\text{sc}_{\l m} = \frac{d}{dr^2} + \frac{f'}{f} \frac{d}{dr} - f^{-2}\left[V_l(r) - \omega_m^2 \right]
\end{align}
with
\begin{align}
	V_\l(r) = f\left[\frac{2M}{r^3} + \frac{\l(\l+1)}{r^2}\right].
\end{align}
The $\mathcal{J}^{(i)}_{\l m}$ in Eq.~\eqref{eq:Lorenz_gauge_FD_ODE} come from the decomposition of the source. When the test body is spinning, $\mathcal{J}^{(i)}_{\l m}$ contains terms proportional to $\delta(r-r_0)$ and $\delta'(r-r_0)$. The $\mathcal{M}$'s are first-order differential operators that couple together the $\hb^{(i)}$'s. Their explicit forms can be found in, e.g., Appendix B of Ref.~\cite{Wardell:2015ada}. For a given $\l m$-mode we have $k_{\l m}$ fields coupled together through the $\mathcal{M}$'s.

The retarded solution to Eq.~\eqref{eq:Lorenz_gauge_FD_ODE} is constructed in two steps. First, the homogeneous solutions are computed by applying retarded boundary conditions then numerically integrating the homogeneous equations. Details on the boundary conditions can be found in, e.g., Ref.~\cite{Akcay:2010dx}. As Eq.~\eqref{eq:Lorenz_gauge_FD_ODE} is a second-order differential equation, the space of homogeneous solutions will be $2k_{\l m}$ dimensional. Let us define the `inner' and `outer' homogeneous solutions by $\hbh^{(i)-}_j$ and $\hbh^{(i)+}_j$, respectively, where $j=1,\dots,k_{\l m}$ indexes the basis of solutions. In this context `inner' means either ingoing radiation and/or regularity at the horizon (and the same with `outer' but at spatial infinity). 

The inhomogeneous solutions to Eq.~\eqref{eq:Lorenz_gauge_FD_ODE} are then computed using the method of variation of parameters. This involves integrating a matrix of homogeneous solutions against the source. The $\delta$- and $\delta'$-functions in the source means that this integration can be done analytically and the inhomogeneous solutions can be written explicitly as
\begin{align}\label{eq:Lorenz_hinhom}
	\hb^{(i)}(r) = \hb^{(i)-}(r) \Theta(r_\p -r) + \hb^{(i)+}\Theta(r-r_\p),\qquad \text{where} \qquad \hb^{(i)\pm}= \sum_{j=1}^k C_{j\p}^\pm \hbh^{(i)\pm}_j(r),
\end{align}
where $\Theta$ is the Heaviside step function and
\begin{align}\label{eq:ret_weighting_coeffs}
	\left(\begin{array}{c} C^-_{j0} \\ C^+_{j0}\end{array}\right) =  \Phi^{-1}(r_0)\left(\begin{array}{c} [\hb'^{(j)}]_{r_\p} \\ {[\hb^{(j)}]_{r_\p}}   \end{array}\right).
\end{align}
Hereafter $[\,\cdot\,]_{r_\p}$ represents the difference in the one-sided limits evaluated at $r_\p$ and $\Phi$ is a matrix of homogeneous solutions given by
\begin{eqnarray}\label{eq:Phi_matrix}
	\arraycolsep=1.4pt\def\arraystretch{1.5}
	\Phi(r) = \left(\begin{array}{c | c}-\hbh^{(i)-}_j & \hbh^{(i)+}_j	\\ \hline -\partial_r \hbh^{(i)-}_j & \partial_r \hbh^{(i)+}_j \end{array}\right).
\end{eqnarray}
In the next section, we discuss how the vector of jump conditions in the right-hand side of Eq.~\eqref{eq:ret_weighting_coeffs} are calculated.

\subsection{Lorenz-gauge source}\label{sec:Lorenz-gauge_source}

We begin by decomposing the source \eqref{eq:FBB_spin_source} into tensor spherical harmonics of the form
\be
	T_{\mu\nu}= \mu \sum_{\l,m}\sum_{i=1}^{10} T^{(i)}_{\l m}(t,r) Y^{(i)\l m}_{\mu\nu}(\theta,\phi;r). \label{eq:T_munu_in_sph_harm}
\ee
The $T^{(i)}_{\l m}(t,r)$ can be computed explicitly using the orthogonality relations of the tensor harmonics \cite{Wardell:2015ada}.
The decomposition for the monopole source is well known \cite{Akcay:2010dx, Akcay:2013wfa} so we focus on the spin-dipole, $\ord(\s)$, term here. 

The standard form for the sources of the frequency-domain Lorenz-gauge field equations \eqref{eq:Lorenz_gauge_FD_ODE} is given by
\be
	\mathcal{J}^{(i)}_{\l m} \equiv  -{16\pi} \f{r}{a^{(i)}f } T^{(i)}_{\l m}, \label{eq:J_i}
\ee
For a spinning body with stress-energy given by Eq.~\eqref{eq:T_munu} the decomposed source is given explicitly by
\begin{align}
	\mathcal{J}^{(i)}_{lm}(r) = -\frac{16\pi \geo{\E}}{f_0^2}\left[(\geo{\alpha}^{(i)} + \s \alpha^{(i)}_\s) \delta(r-r_\p) + \sigma \beta_\s^{(i)} \delta'(r-r_\p) \right]\left\{\begin{array}{c} Y^{\ell m*}(\pi/2,\O t),\;\;i=1,\dots,7\\ Y^{\ell m*}_{,\theta}(\pi/2,\O t),\;\;i=8,9,10 \end{array}\right.,
\end{align}
where the geodesic terms are given by
\begin{align}
  \geo{\alpha}^{(1)} = f_\p^2/r_\p, \quad
  \geo{\alpha}^{(2)} = 0, \quad
  \geo{\alpha}^{(3)} = f_\p/r_\p, \quad
  \geo{\alpha}^{(4)} = 2i m f_\p \Og, \quad
  \geo{\alpha}^{(5)} = 0, \quad
  \nonumber \\
  \geo{\alpha}^{(6)} = r_\p\Og^2, \quad
  \geo{\alpha}^{(7)} = r_\p\Og^2\left[\ell(\ell+1) - 2 m^2\right], \quad
  \geo{\alpha}^{(8)} = 2f_\p\Og, \quad
  \geo{\alpha}^{(9)} = 0, \quad
  \geo{\alpha}^{(10)} = 2imr_\p\Og^2.
\end{align}
These coefficients are the same as those given in Eq.~(B12) of Ref.~\cite{Wardell:2015ada}. When the secondary is spinning the additional terms are given by
\begin{align}
	\alpha^{(1)}_\s =  f_\p [f_\p r_\p \utsbar +2M (4M - r_\p)\Og]/r_\p^2 , \quad
	\alpha^{(2)}_\s = -i m M f_\p(f_\p - r_\p^2\Og^2)/r_\p^2,	\quad
	\alpha^{(3)}_\s = f_\p \utsbar/r_\p,	\quad
	\nonumber \\
	\alpha^{(4)}_\s = 2 i m \Og(f_\p r_\p \utsbar + M^2\Og)/r_\p,	\quad
	\alpha^{(5)}_\s = m^2 M \Og (r_\p - 3M)/r_\p^2,					\quad
	\alpha^{(6)}_\s =  r_\p \utsbar \Og^2,						\quad
	\nonumber \\
	\alpha^{(7)}_\s = [l(l+1) - 2m^2]r_\p \utsbar \Og^2,	\quad
	\alpha^{(8)}_\s = \alpha^{(4)}_\s/(im),	\quad
	\alpha^{(9)}_\s = i m M \Og (3M-r_\p)/r_\p^2	\quad
	\alpha^{(10)}_\s = 2 i m M r_\p \utsbar \Og^2,
\end{align}
where $\utsbar = u^t_\s/\geo{u}^t$ and
\begin{align}
	\beta^{(1)}_\s = -M f_\p^2 \Og, 	\quad
	\beta^{(2)}_\s = 0,			\quad
	\beta^{(3)}_\s = -M f_\p \Og	,	\quad
	\beta^{(4)}_\s = - i m M f_\p (r_\p - M)/r_\p^2,			\quad
	\beta^{(5)}_\s = 0,
	\nonumber \\
	\beta^{(6)}_\s = -M f_\p\Og,		\quad
	\beta^{(7)}_\s = -M f_\p(l(l+1) - 2m^2)\Og,	\quad
	\beta^{(8)}_\s = \beta^{(4)}/(im),	\quad
	\beta^{(9)}_\s = 0,	\quad
	\beta^{(10)}_\s = - 2 i m f_\p \Og.
\end{align}
As a check on these sources, we have explicitly verified that they are divergence free at the mode level.

To compute the junction conditions $[\,\cdot\,]_{r_\p}$ required in Eq.~\eqref{eq:ret_weighting_coeffs} we substitute Eq.~\eqref{eq:Lorenz_hinhom} into the radial equation \eqref{eq:Lorenz_gauge_FD_ODE}. Matching coefficients of the $\delta$'s and $\delta'$'s we arrive at
\begin{align}\label{eq:Lorenz_gauge_junction_conditions}
	[\hb^{(j)}]_{r_\p}  &= -\frac{16\pi \geo{\E}\s}{f_0^2}\beta_\s^{(j)},				\\
	[\hb'^{(j)}]_{r_\p}	&= -\frac{16\pi \geo{\E}}{f_0^2}\left[\geo{\alpha}^{(j)} + \s \left( \alpha_\s^{(j)} + \tensor{\mathcal{N}}{^{(j)}_{\!\!(k)}}\beta_\s^{(k)}  \right)\right]. 
\end{align}
The $\tensor{\mathcal{N}}{^{(j)}_{\!\!(k)}}$ come from the first order derivatives that appear in the $\tensor{\mathcal{M}}{^{(j)}_{\!\!(k)}}$ for the $i=1,2,4,8$ fields. Curiously the contributions from $i=2$ cancel out leaving the only non-zero contributions as
\begin{subequations}
\begin{align}
	\tensor{\mathcal{N}}{^{(1)}_{\!\!(k)}}\beta_\s^{(k)} &= 4M\beta^{(3)}_\s /r_\p^2,									\\
	\tensor{\mathcal{N}}{^{(4)}_{\!\!(k)}}\beta_\s^{(k)} &= \frac{2M\beta^{(4)}_\s}{(r_\p - 2M)r_\p} ,			    	\\
	\tensor{\mathcal{N}}{^{(8)}_{\!\!(k)}}\beta_\s^{(k)} &= \frac{2M\beta^{(8)}_\s}{(r_\p - 2M)r_\p} .
\end{align}
\end{subequations}
The junction conditions \eqref{eq:Lorenz_gauge_junction_conditions} tell us that in the non-spinning case ($\s=0$) the modes of the retarded field will be continuous at the particle with a discontinuity in some of their their derivatives. In the spinning case, both the fields and their derivative can be discontinuous at the particle.

\subsection{Numerical calculation}

As this work is only concerned with the radiated flux and the dissipative local force, we are not required to calculate any static ($\omega=0$) modes. 
For circular orbits, this translates to calculating modes with $ l\ge 2$ and $m \neq 0$. 
Unlike for the static modes, the field equations for the radiative modes do not admit closed-form analytic solutions. 
In the Lorenz gauge case, we opt to solve for these modes by numerically integrating the field equations. 
Our procedure for this follows closely to Ref.~\cite{Wardell:2015ada}.

\begin{figure}
	\includegraphics[width=8.5cm]{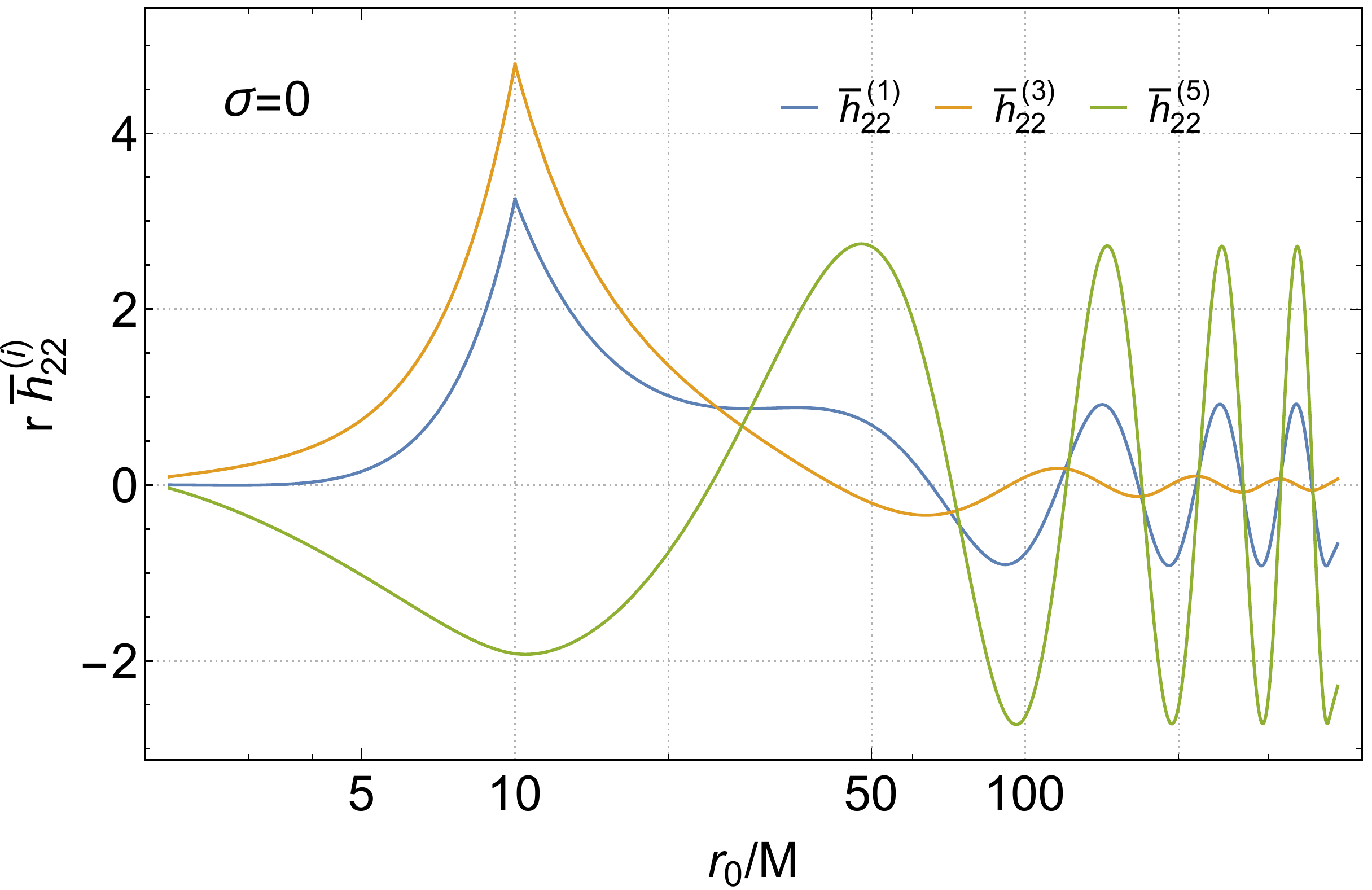}\quad
	\includegraphics[width=8.5cm]{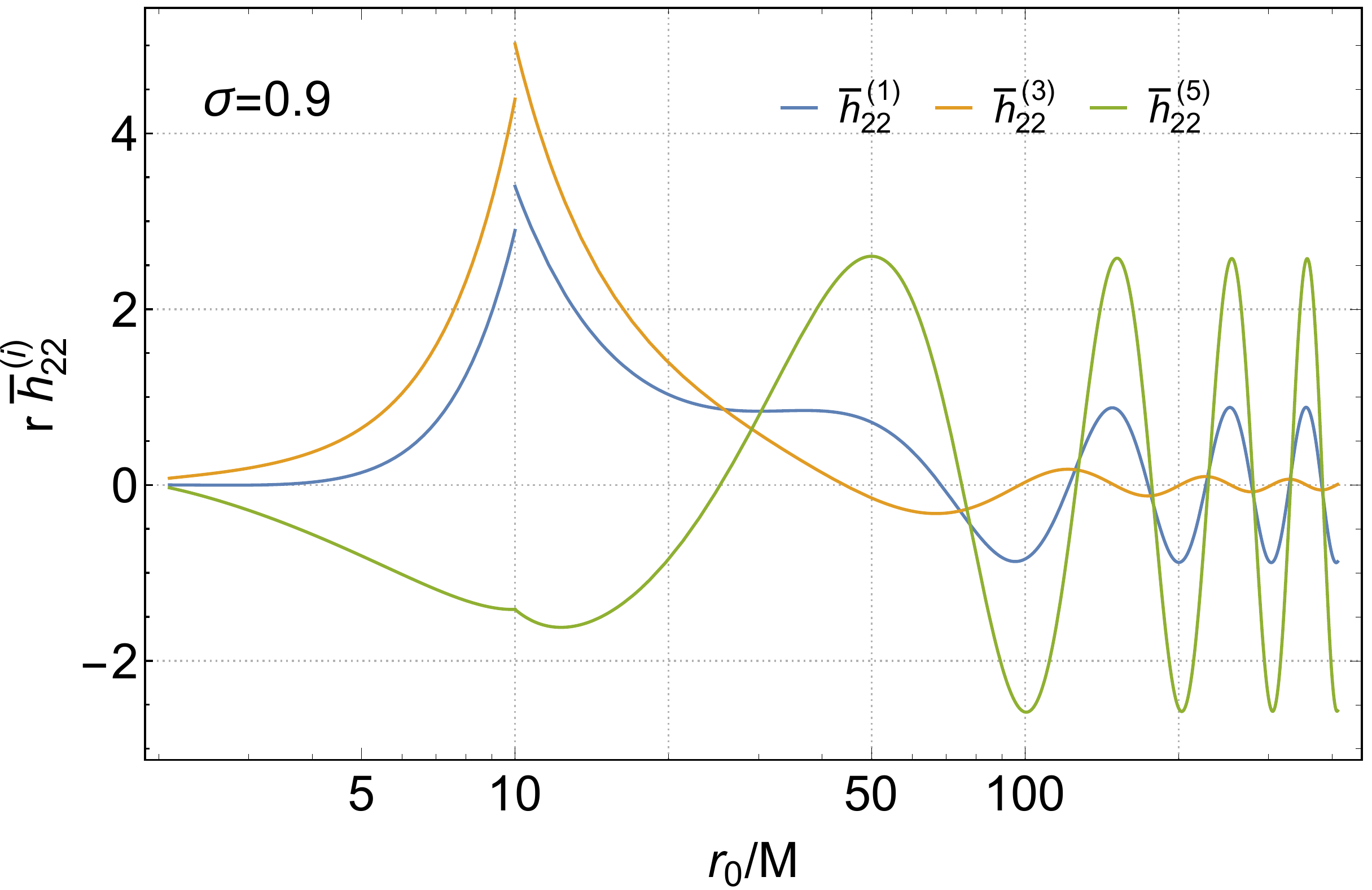}
	\caption{
	The $\l=2, m=2$ mode of the Lorenz gauge metric perturbation for a particle orbiting at a radius of $r_0=10\M$ with no spin (left panel) and with spin magnitude $\s = 0.9$ (right panel). In the non-spinning case the metric perturbation is continuous at the a particle whereas for the spinning case some of the fields of the metric perturbation are discontinuous. For this  mode, we see that the $i=1$ and $i=3$ fields are discontinuous whilst the $i=5$ remains continuous. To avoid cluttering the figure we have not shown the $i=\{2,4,6,7\}$ fields, which exhibit similar behaviour to the ones shown.}\label{fig:Lorenz_metric_pert}
\end{figure}

For each $lm$-mode we begin by solving for the homogeneous solutions by applying appropriate boundary conditions following Ref.~\cite{Akcay:2010dx}. 
We then numerically integrate the field equations in \texttt{Mathematica} using the \texttt{NDSolve[]} function. 
These two steps differ from previous work only by the mode frequency which is now given by $\omega = m\O$ rather than $\omega = m\Og$. 
This gives us a basis of homogeneous solutions, $\hbh^{(i)\pm}_j$, that span the solution space.
The inhomogeneous solutions are then constructed via Eq.~\eqref{eq:Lorenz_hinhom}. 
The radiated flux is then computed directly from the weighting coefficients, $C^\pm_{j0}$'s (see formula in Ref.~\cite{Akcay:2010dx}). 
The metric perturbation is constructed from the $\hb^{(i)}$'s using the formula in appendix A.6 of Ref.~\cite{Wardell:2015ada}. 
We give an example of the metric perturbation for the $\l=2,m=2$ mode in Fig.~\ref{fig:Lorenz_metric_pert}. 
From the values of the metric perturbation and its derivatives at the particle we then compute the change to the local energy using Eq.~\eqref{eq:SimpleSpinFlux}.

\section{Explicit results for the flux and local forces} \label{Sec:Results}

In this section we given explicit analytic PN as well as numerical results for the radiated fluxes and the rate of change of the local energy. Using these results, we verify that the flux balance law given in Eq.~\eqref{eq:SummaryEquation} holds through $\mathcal{O}(\s)$.

\subsection{Analytical post-Newtonian results}

We include in our calculation modes with $\ell\leq6$. This results in expressions for the asymptotic fluxes and the local forcings valid to 5.5PN order.

When working in the radiation gauge, as we do for the PN calculation, we find the interesting behaviour that both of the terms appearing in the right hand side of \eqref{eq:SummaryEquation} are, at $\mathcal{O}(\s)$, divergent in $\ell$ and discontinuous when taking the radial limit to the position of the particle from the left and from the right. Explicitly we see that in the large-$\ell$ limit
\begin{subequations}
\begin{align}
	\left(S^{\gamma \delta} u^\beta \nabla^{(0)}_\delta \mathcal{L}_{\xi} h_{\gamma \beta}\right)^+&=
	\frac{6(\ell+1)}{\ell-1}\s y^{11/2} +\frac{3 \left(27 \ell ^3+40 \ell ^2-8 \ell -9\right)}{(\ell -1) (2 \ell -1) (2 \ell +3)}\s y^{13/2}+\mathcal{O}(y^7),\\
	\left(S^{\gamma \delta} u^\beta \nabla^{(0)}_\delta \mathcal{L}_{\xi} h_{\gamma \beta}\right)^-&=
	\frac{6\ell}{2+\ell}\s y^{11/2} +\frac{3 \left(27 \ell ^3+41 \ell ^2-7 \ell -12\right)}{(\ell +2) (2 \ell -1) (2 \ell +3)}\s y^{13/2}+\mathcal{O}(y^7),\\
	\left(u^\alpha u^\beta \mathcal{L}_{\xi} h_{\alpha \beta}\right)^+ &=\frac{6(\ell+1)}{\ell-1}\s y^{11/2} +\frac{3 \left(27 \ell ^3+40 \ell ^2-8 \ell -9\right)}{(\ell -1) (2 \ell -1) (2 \ell +3)}\s y^{13/2}+\mathcal{O}(y^7),\\
	\left(u^\alpha u^\beta \mathcal{L}_{\xi} h_{\alpha \beta}\right)^- &=\frac{6\ell}{2+\ell}\s y^{11/2} +\frac{3 \left(27 \ell ^3+41 \ell ^2-7 \ell -12\right)}{(\ell +2) (2 \ell -1) (2 \ell +3)}\s y^{13/2}+\mathcal{O}(y^7).
\end{align}
\end{subequations}
Given the relative sign difference between these two terms in \eqref{eq:SummaryEquation} we see that all polynomial divergence is removed in the combination. For the low-$\ell$ values the two terms are explicitly \textit{not} equal. For example, at $\ell=2$:
\begin{subequations}
\begin{align}
	\left(S^{\gamma \delta} u^\beta \nabla^{(0)}_\delta \mathcal{L}_{\xi} h_{\gamma \beta}\right)^+
	&=18 \s y^{11/2}+\frac{1139}{35}\s y^{13/2}+\mathcal{O}(y^7),\\
	\left(S^{\gamma \delta} u^\beta \nabla^{(0)}_\delta \mathcal{L}_{\xi} h_{\gamma \beta}\right)^-
	&=3 \s y^{11/2}-\frac{347}{70}\s y^{13/2}+\mathcal{O}(y^7),\\
	\left(u^\alpha u^\beta \mathcal{L}_{\xi} h_{\alpha \beta}\right)^+ 
	&=\frac{64}{5}y^5+18\s y^{11/2}-\frac{14384}{315}y^6+\left(\frac{256 \pi }{5}+\frac{579}{35}\s\right)+ y^{13/2}+\mathcal{O}(y^7) ,\\
	\left(u^\alpha u^\beta \mathcal{L}_{\xi} h_{\alpha \beta}\right)^-
	&=\frac{64}{5}y^5+3\s y^{11/2}-\frac{14384}{315}y^6+\left(\frac{256 \pi }{5}-\frac{1467}{70}\s\right)+ y^{13/2}+\mathcal{O}(y^7).
\end{align}
\end{subequations}
This verifies that, as one would expect for a purely radiative quantity, the mode sum of the dissipated energy is exponentially convergent in $\l$. This delicate cancellation of polynomial behaviour in $\ell$ serves as a useful consistency check within the calculation.

Combining these expressions and computing the sum over $\l$-modes we arrive at our 5.5PN accurate expression which is identical to that calculated from the asymptotic Teukolsky fluxes:
\begin{align}
	\frac{1}{u^t}\frac{D\mathcal{E}}{d \tau}=&\frac{32}{5}y^5\Bigg[1-\frac{1247 }{336}y+\left(4 \pi -\frac{5 }{4}\s\right) y^{3/2}-\frac{44711 }{9072}y^2+\left(-\frac{8191 
   }{672}\pi-\frac{13 }{16}\s\right) y^{5/2}+\bigg(\frac{6643739519}{69854400}-\frac{1712 }{105} \gamma+\frac{16 }{3}\pi
   ^2 \nn \\ 
   &-\frac{3424 }{105}\log (2)-\frac{856}{105} \log (y)-\frac{31  }{6}\pi \s\bigg) y^3
   +\left(-\frac{16285 
   }{504}\pi+\frac{9535 }{336}\s\right) y^{7/2}
   +\bigg(-\frac{319927174267}{3178375200}+\frac{232597 
   }{4410} \gamma\nn \\
   &-\frac{1369 }{126}\pi ^2+\frac{39931 }{294}\log (2)-\frac{47385 }{1568}\log
   (3)+\frac{232597 }{8820}\log (y)-\frac{7163  }{672}\pi \s\bigg) y^4
   +\left\{\frac{265978667519 }{745113600} \pi-\frac{6848 
   }{105}\gamma  \pi\right. \nn \\
   &\left.-\frac{13696}{105} \pi  \log (2)-\frac{3424}{105} \pi  \log (y)+ \left(-\frac{37454731}{453600}+\frac{107
  }{5} \gamma -7 \pi ^2+\frac{13589}{315} \log (2)+\frac{107 }{10}\log (y)\right)\s\right\}
   y^{9/2} \nn \\
    &+\bigg(-\frac{32866400674911451}{36815119941600}+\frac{916628467}{7858620} \gamma -\frac{424223 }{6804}\pi
   ^2-\frac{83217611 }{1122660}\log (2)+\frac{47385}{196} \log (3) \nn \\
   &+\frac{916628467
  }{15717240} \log (y)+\frac{384707}{3024} \pi  \s\bigg) y^5
   +\left\{\frac{8399309750401  }{101708006400}\pi+\frac{177293
   }{1176} \gamma  \pi+\frac{8521283 }{17640}\pi  \log (2)-\frac{142155}{784} \pi  \log (3)\right. \nn \\
   &\left.+\frac{177293 }{2352}\pi  \log (y)+
   \left(-\frac{2227389947}{3880800}+\frac{211 }{6} \gamma-30 \pi ^2-\frac{96179}{4410} \log (2)+\frac{142155 }{1568}\log
   (3)+\frac{211}{12} \log (y)\right)\s\right\} y^{11/2}\bigg]\nn\\
   &+\mathcal{O}(y^{11}).
	\end{align}
This PN series can be found digitally in the \texttt{PostNewtonianSelfForce} package of the Black Hole Perturbation Toolkit \cite{BHPToolkit}. We note also that as extra verification the linear in $\s$ terms here agree with the leading order in the mass-ratio terms of the flux expansions from post-Newtonian theory, e.g. Eq.~(414) of \cite{Blanchet:2013haa}.

\subsection{Numerical results}

An important feature of our numerical results is that, despite the orbital dynamics and the perturbation source being linear in $\s$, the calculated flux and local forces are not. 
This occurs because our calculation contains products of terms that have been linearized in $\s$ as well as products with a term that is quadratic in $\s$. 
The latter comes from the calculation of the homogeneous solutions to the Teukolsky or Lorenz-gauge equations which both have an $\omega^2 = (m\Omega)^2$ term in their potentials. 
In principle, one could expand the field equations to leading order in $\s$, write down new boundary conditions and solve for the linear in $\s$ piece of the homogeneous solutions. 
With this, one would have all the terms in the calculation expanded to leading order in $\s$ and could then carefully ensure that only the linear terms were retained when any products of these terms were taken. 
We have not attempted to do this in this work. 
Instead, at each orbital radius we compute the fluxes and change to the local energy for a range of values of $\sigma$, fit the results to polynomial in $\sigma$ and extract the linear in $\sigma$ piece. 
With this approach we observe in our results that the agreement between the Teukolsky and Lorenz-gauge calculations holds to high precision through $\mathcal{O}(\s)$ -- see Table \ref{table:numerical_results} for details.
We do not find this observation holds for the higher order in $\s$ terms but then we would not expect them to.

In fitting for the linear-in-$\s$ piece of the result, we compute the fluxes and local forces for $\sigma=\{0, \pm 0.1, \pm 0.2, \pm 0.3,$ $ \pm 0.5, \pm 0.7, \pm 0.9\}$. 
We then perform a least-squares fit to a tenth-order polynormal in $\s$ and extract up to the linear-in-$\s$ piece.\footnote{The order of the polynomial in this fit may seem high but note that the solutions to the Teukolsky equation have contributions up to least $\mathcal{O}(\s^3)$ as the homogeneous equations are $\mathcal{O}(\s^2)$ via the $\omega^2$ in the field equations and the Teukolsky source is $\mathcal{O}(\s)$. The metric reconstruction procedure then introduces many additional powers of $\s$. In principle we could linearize the metric reconstruction formula with respect to $\s$ but instead, as our data is of high quality, we find it easier to perform a high-order fit.}
The need to compute data for many different values of $\s$ at each orbital radius adds greatly to the computation burden of the calculation. 
Fortunately, circular orbit calculations in the frequency domain are sufficiently fast that this is not a problem. 
For, e.g., eccentric orbits the cost of the frequency domain calculation rises rapidly as the eccentricity of the orbit increases \cite{Osburn:2014hoa}.
In this case, the additional computational cost of repeating the calculation for many values of $\s$ is likely to be too burdensome and fully linearising the calculation in $\s$ as outlined above would be advantageous. 

\begin{figure}
	\includegraphics[width=8.5cm]{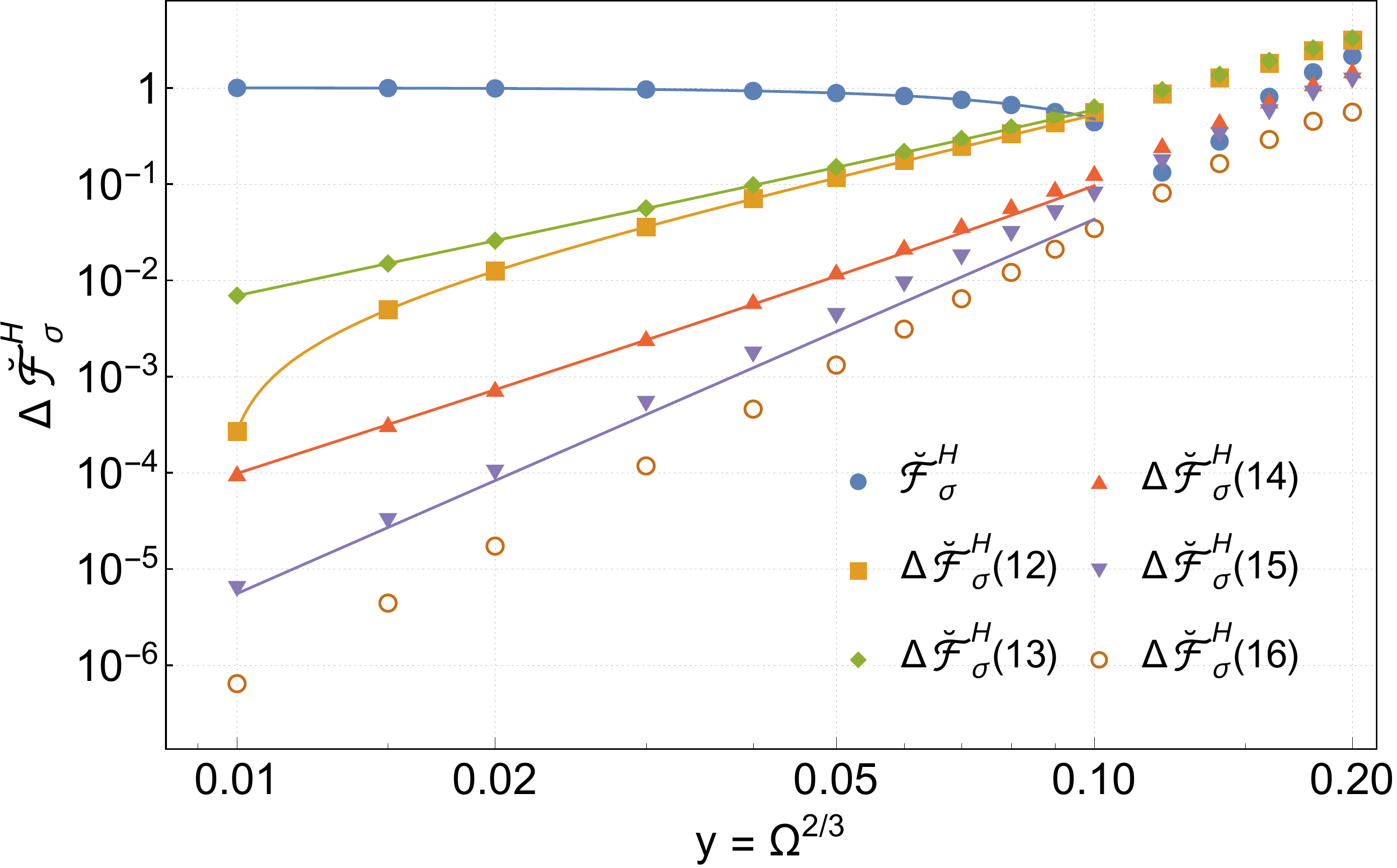}\qquad
	\includegraphics[width=8.5cm]{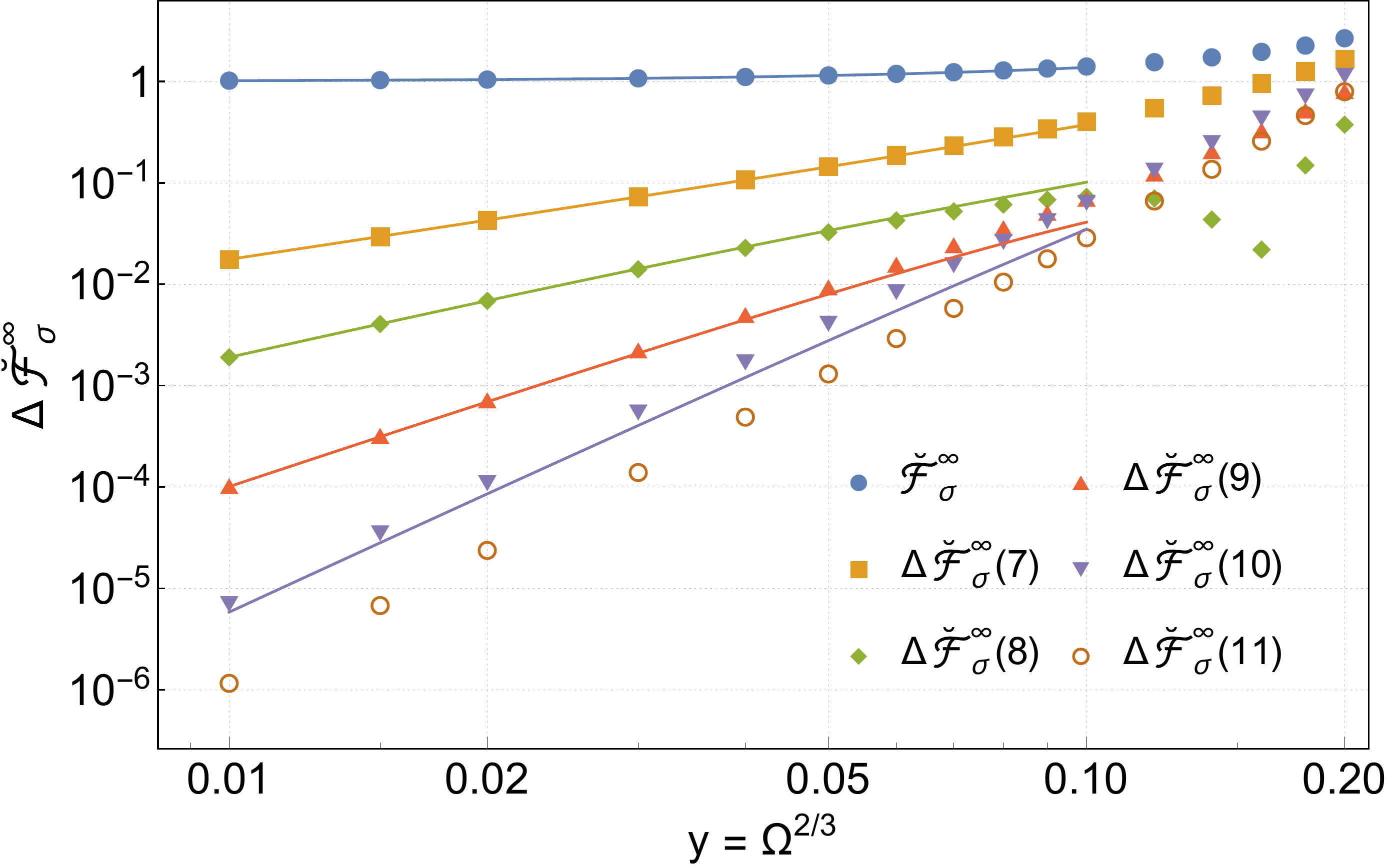}
	\caption{
		Comparison between the (Lorenz-gauge) numerical data and the post-Newtonian series for the linear-in-$\s$ contribution to the flux.
		Left panel: Difference between the numerical and PN results for the horizon flux. 
		In this panel we define the normalized horizon flux by $\fluxbH_\s \equiv \fluxH_\s / (-96/5 y^{23/2})$.
		We also define $\fluxbHPN_\s(n)$ as the (normalized) PN series for the horizon flux truncated at $\mathcal{O}(y^n)$ and plot these as solid curves for $y \le 0.1$.
		The difference between the numerical and PN results is then given by $\Delta \fluxbH_\s(n) \equiv |\fluxbH_\s - \fluxbHPN_\s(n)|$.
		As we subtract ever higher order PN series from the numerical data we see that the residual drops in amplitude.
		This cross check on our numerical and analytic results gives us confidence that both are correct.
		Right panel: The same as the left panel but for the infinity flux. In this panel the normalized infinity flux is given by $\fluxbI_\s \equiv \fluxI_\s / (-8 y^{13/2})$.
	}\label{fig:PN_comparison}
\end{figure}

Using the fitting method described above we can separate the spinning and non-spinning contributions to the energy flux in the form
\begin{align}
	\flux(r_0) = \geo{\flux}(r_0) + \s \flux_\s(r_0)
\end{align}
We will also define $\langle \cdot\rangle_\s$ as an operator that extracts the $\mathcal{O}(\s)$ piece of a quantity, e.g., $\langle \flux\rangle_\s = \flux_\s$. 
We can further separate the flux into the piece radiated to infinity and the piece radiated down the horizon. 
Concentrating on the $\mathcal{O}{(\s)}$ piece we write\begin{align}
	\flux_\s(r_0) = \fluxH_\s(r_0) + \fluxI_\s(r_0)
\end{align}
We give results for $\fluxH_\s(r_0)$, $\fluxI_\s(r_0)$, and the rate of change of the local energy in Table~\ref{table:numerical_results}. We also give the same quantities computed at fixed $y$ in Table~\ref{table:numerical_results_fixed_y}. In all cases we find excellent agreement between the asymptotic fluxes and the local change in the energy, as indicated by the fifth column in the tables. We also compare our numerical results with our PN series and find excellent agreement -- see Fig.~\ref{fig:PN_comparison}. As a further check, we have compared our data for $\fluxI_\s$ against the results from the time-domain Teukolsky code presented in Ref.~\cite{Harms:2015ixa}. That comparison is presented in Ref.~\cite{Nagar:2019wrt} where we found agreement to within the $\sim1\%$ level errors bars on the time-domain results.

For all the orbital radii we have explored we find that the flux decreases for a spin-aligned binary (with respect to a non-spinning binary). This decrease in the flux will lead to spin-aligned binaries taking longer to inspiral than non-spinning binaries. It is interesting to note that this is consistent with the ``orbital hangup'' effect observed in numerical relativity simulations \cite{Campanelli:2006uy}.

\begin{table}[]
\begin{tabular}{l | l l l  | l | l}
$r_0$ & $\geo{\flux}$  & $\fluxH_\s$ & $\fluxI_\s$  & $\langle d\mathcal{E}/d\tau\rangle_\s$   & $\Delta^\text{rel} $\\
\hline
$6$		& $9.4033935628\times 10^{-4}$	& $-2.4411027706\times 10^{-6}$		& $-5.050521990\times 10^{-4}$		& $-7.6294600853\times 10^{-4}$		& $2.7\times 10^{-11}$	\\
$8$		& $1.9610454858\times 10^{-4}$	& $-5.8512615270699\times 10^{-8}$	& $-6.2795524582\times 10^{-5}$		& $-8.2793540332\times 10^{-5}$		& $5.1\times 10^{-13}$	\\
$10$	& $6.1516316785\times 10^{-5}$	& $-4.02409747536897\times 10^{-9}$	& $-1.3528384048576\times 10^{-5}$	& $-1.66725567034\times 10^{-5}$	& $6.9\times 10^{-13}$	\\
$12$	& $2.4291700945\times 10^{-5}$	& $-4.917303952656\times 10^{-10}$	& $-3.967615345444\times 10^{-6}$	& $-4.694436955265\times 10^{-6}$	& $4.7\times 10^{-13}$	\\
$20$	& $1.8714709114\times 10^{-6}$	& $-1.7044774934187\times 10^{-12}$	& $-1.363681646442\times 10^{-7}$	& $-1.499163835028\times 10^{-7}$	& $1.0\times 10^{-13}$	\\
$30$	& $2.4864755005\times 10^{-7}$	& $-2.144634376248\times 10^{-14}$	& $-9.6955394911065\times 10^{-9}$	& $-1.03086338505\times 10^{-8}$	& $4.1\times 10^{-13}$	\\
$40$	& $5.9501545594\times 10^{-8}$	& $-9.927811950102\times 10^{-16}$	& $-1.49558022978768\times 10^{-9}$	& $-1.56494549168\times 10^{-9}$	& $2.3\times 10^{-12}$	\\
$50$	& $1.9624578561\times 10^{-8}$	& $-9.25922620716\times 10^{-17}$	& $-3.51467899595\times 10^{-10}$	& $-3.64338707066\times 10^{-10}$	& $1.5\times 10^{-12}$	\\
$60$	& $7.9264448530\times 10^{-9}$	& $-1.33975153331\times 10^{-17}$	& $-1.07706168184\times 10^{-10}$	& $-1.1096468581\times 10^{-10}$	& $3.5\times 10^{-11}$	\\
$70$	& $3.6818812737\times 10^{-9}$	& $-2.620714098344\times 10^{-18}$	& $-3.963027373213\times 10^{-11}$	& $-4.0651669377\times 10^{-11}$	& $3.3\times 10^{-12}$	\\
$80$	& $1.8945359109\times 10^{-9}$	& $-6.38761880534\times 10^{-19}$	& $-1.66688751664\times 10^{-11}$	& $-1.7043065115\times 10^{-11}$	& $2.9\times 10^{-11}$	\\
$90$	& $1.0541122976\times 10^{-9}$	& $-1.84096376783\times 10^{-19}$	& $-7.7649000465\times 10^{-12}$	& $-7.919293126\times 10^{-12}$		& $1.3\times 10^{-11}$	\\
$100$	& $6.2382034734\times 10^{-10}$	& $-6.05434134454\times 10^{-20}$	& $-3.92050069646\times 10^{-12}$	& $-3.9904601554\times 10^{-12}$	& $5.4\times 10^{-12}$	\\
\end{tabular}
\caption{
	Contribution to the radiated flux and rate of change of the local energy for a spinning body moving on a circular orbit of radius $r_0$ about a Schwarzschild black hole.
	All the data in this table has been adimensionalzied such that, e.g., $\fluxH_\s \equiv [M^2/(\mu^2\sigma)]\fluxH_\s$.
	The flux results, presented in the second through forth columns, are made with both a Teukolsky and a Lorenz gauge code. 
	In these columns we present all the digits that agree between these two codes. 
	The second column shows the geodesic ($\s=0$) results for the total flux (horizon plus infinity).
	As these are presented elsewhere in the literature \cite{Akcay:2010dx} we truncate the data in this column at 11 significant digits.
	The third and forth columns give the $\mathcal{O}(\s)$ contribution to the horizon and infinity flux, respectively.
	For the local force we find excellent agreement between the radiation and Lorenz gauge results to a relative error of better than $10^{-8}$.
	This lower precision (relative to the flux) comes from the complicated metric reconstruction into the radiation gauge.
	Thus we instead show results from the Lorenz-gauge code which works with extended precision, truncating the result based how well the flux balance formula is satisfied.
	The fifth column shows the $\mathcal{O}(\s)$ contribution to the local force.
	The final column shows the relative difference $\Delta^\text{rel} \equiv |1 - \langle\flux u^t\rangle_\s/\langle d\mathcal{E}/d\tau\rangle_\s|$. 
	This difference is always less than $4\times10^{-11}$ which shows the excellent numerical agreement we find using the flux balance law.
	Orbits with $r_0 \le 20$ were computed with $l_\text{max} = 20$ which is our truncation value for
	the $l$-mode sums in Eqs.~(\ref{eq:hb_decomp}, \ref{eq:T_munu_in_sph_harm}). 
	All other orbits were computed with $l_\text{max}=15$. 
	The data in the second through fifth columns can be found digitally in the Black Hole Perturbation Toolkit \cite{BHPToolkit}.
}
\label{table:numerical_results}
\end{table}

\begin{table}[]
\begin{tabular}{l | l l l  | l | l}
$y$ 	& $\geo{\flux}$  & $\fluxH_\s$ & $\fluxI_\s$  & $\langle d\mathcal{E}/d\tau\rangle_\s$   & $\Delta^\text{rel} $\\
\hline
$0.2$	&	$2.79273701868\times 10^{-3}$	&	$3.77193403191\times 10^{-7}$	&	$-6.104060211\times 10^{-4}$	&	$-9.64540266941\times 10^{-4}$	&	$3.0\times 10^{-13}$	 \\
$0.18$	&	$1.46844806236\times 10^{-3}$	&	$7.605414762924\times 10^{-8}$	&	$-2.60585846715\times 10^{-4}$	&	$-3.841007341364\times 10^{-4}$	&	$6.5\times 10^{-14}$	 \\
$0.16$	&	$7.467542778218\times 10^{-4}$	&	$1.089805069009\times 10^{-8}$	&	$-1.050643019744\times 10^{-4}$	&	$-1.456828594266\times 10^{-4}$	&	$1.4\times 10^{-14}$	 \\
$0.14$	&	$3.5876589417\times 10^{-4}$	&	$8.0692632306\times 10^{-10}$	&	$-3.8940747125\times 10^{-5}$	&	$-5.1130646432\times 10^{-5}$	&	$7.5\times 10^{-12}$	 \\
$0.12$	&	$1.582281533\times 10^{-4}$		&	$-6.539052356\times 10^{-11}$	&	$-1.280679512\times 10^{-5}$	&	$-1.600857564\times 10^{-5}$	&	$9.9\times 10^{-11}$	 \\
$0.1$	&	$6.151631678\times 10^{-5}$		&	$-2.669935713\times 10^{-11}$	&	$-3.549175593\times 10^{-6}$	&	$-4.242108121\times 10^{-6}$	&	$3.4\times 10^{-11}$	 \\
$0.09$	&	$3.590633623\times 10^{-5}$		&	$-1.014769938\times 10^{-11}$	&	$-1.710319876\times 10^{-6}$	&	$-2.001789881\times 10^{-6}$	&	$1.5\times 10^{-11}$	 \\
$0.08$	&	$1.9757908533\times 10^{-5}$	&	$-3.1009617821\times 10^{-12}$	&	$-7.6206608517\times 10^{-7}$	&	$-8.7415330798\times 10^{-7}$	&	$6.0\times 10^{-12}$	 \\
$0.07$	&	$1.0079767299\times 10^{-5}$	&	$-7.5507222921\times 10^{-13}$	&	$-3.0721180533\times 10^{-7}$	&	$-3.4564113472\times 10^{-7}$	&	$2.0\times 10^{-12}$	 \\
$0.06$	&	$4.6528705441\times 10^{-6}$	&	$-1.4058811966\times 10^{-13}$	&	$-1.0855179435\times 10^{-7}$	&	$-1.1987555833\times 10^{-7}$	&	$1.1\times 10^{-12}$	 \\
$0.05$	&	$1.8714709114\times 10^{-6}$	&	$-1.8506079813\times 10^{-14}$	&	$-3.2008999168\times 10^{-8}$	&	$-3.4718654292\times 10^{-8}$	&	$1.2\times 10^{-12}$	 \\
$0.04$	&	$6.1579196033\times 10^{-7}$	&	$-1.4966312714\times 10^{-15}$	&	$-7.255453657\times 10^{-9}$	&	$-7.7343411813\times 10^{-9}$	&	$1.9\times 10^{-12}$	 \\
$0.03$	&	$1.47265886605\times 10^{-7}$	&	$-5.67900033301\times 10^{-17}$	&	$-1.08380957\times 10^{-9}$		&	$-1.13614119765\times 10^{-9}$	&	$5.5\times 10^{-13}$	 \\
$0.02$	&	$1.9624578561\times 10^{-8}$	&	$-5.4913567205\times 10^{-19}$	&	$-7.5512423521\times 10^{-11}$	&	$-7.7885118542\times 10^{-11}$	&	$4.7\times 10^{-12}$	 \\
$0.015$	&	$4.6933548927\times 10^{-9}$	&	$-2.0239012136\times 10^{-20}$	&	$-1.1490337069\times 10^{-11}$	&	$-1.1757935781\times 10^{-11}$	&	$1.9\times 10^{-12}$	 \\
$0.01$	&	$6.238203473\times 10^{-10}$	&	$-1.91947959\times 10^{-22}$	&	$-8.140678916\times 10^{-13}$	&	$-8.265607122\times 10^{-13}$	&	$7.3\times 10^{-11}$	 \\
\end{tabular}
\caption{
	The same as a Table \ref{table:numerical_results} but computed at fixed values of $y$. 
	This data is used to make the comparison with the PN series presented in Fig.~\ref{fig:PN_comparison}. 
	All the data in this table is computed using our Lorenz-gauge code. 
	All digits shown are accurate with the error bars being set by the difference in between the left-hand and right-hand sides of the flux balance formula in Eq.~\eqref{eq:SummaryEquation} (this difference is given in the final column). 
	All the data in this table can be found digitally in the Black Hole Perturbation Toolkit \cite{BHPToolkit}. 
	Orbits with $\{y \ge 0.16, 0.05 \le y < 0.16, y < 0.05 \}$ were computed with $l_\text{max} = \{30, 20, 15\}$, respectively.}
\label{table:numerical_results_fixed_y}
\end{table}

\section{Conclusion} \label{Sec:Conclusions}

In order to produce a post-adiabatic waveform for EMRIs it is crucial to include the effects of the spin of the secondary. Formally these effects enter the waveform phase at the same order as first-order conservative and second-order dissipative self-force effects. No EMRI inspiral and waveform model is complete at post-adiabatic order without including all these contributions. Whilst the self-force contributions have received considerable attention, the equally-as-important spinning contributions have not. This paper represents a significant step forward in addressing these spinning contributions in complementary ways. 

On a formal level, we provide a clear mapping from the `easy to compute’  asymptotic fluxes to the local evolution of the energy and angular momentum. 
These relations are valid for arbitrary spin and non-resonant orbital configurations in both Schwarzschild and Kerr spacetime. 
Since the relation between the four-velocity of a spinning particle and its quasi-conserved energy depends on the spin tensor, knowledge of these asymptotic fluxes does not completely allow one to compute an inspiral. 
It is also necessary to integrate the evolution equation for the spin tensor. Notably this equation only depends on the first order metric perturbation, i.e., it is independent of the dipolar contribution to the stress energy tensor. 
Thus, while our flux balance law does not completely determine the evolution, it completely replaces the computation of the local metric perturbation sourced by the dipolar stress energy with the computation of the asymptotic amplitudes of the Teukolsky functions.

On a computational level, we have developed codes which calculate both the local metric perturbation and the asymptotic fluxes to linear order in the spin of the small body. These codes are in two different gauges: radiation and Lorenz. 
We developed two radiation gauge codes, one analytic in the form of a post-Newtonian series, and the other numerical. 
The Lorenz gauge code is numerical. 
The main result from these codes for this paper is an explicit validation of the energy flux-balance law for a spinning particle in Schwarzschild spacetime on a circular orbit, with its spin vector aligned with the orbital angular momentum. More generally, these codes provide the foundation for the much more generic codes which will be needed to drive more complicated orbital and spin configurations. 

There are a large number of ways in which the work of this paper can be applied or extended. We give these below (in no particular order):
\begin{enumerate}[label=(\roman*)]
\item Omitted from this work is a derivation of a flux balance law for the Carter constant. We expect that the methods used here should be applicable to relate the Carter constant evolution to asymptotic quantities in a similar manner to the non-spinning case.

\item Our expressions are not valid for cases of orbital resonance. 
A recent work by Isoyama et al.~\cite{Isoyama:2018sib} successfully derived flux balance expressions for the non-spinning case which are valid during orbital resonance. 
While the calculations are more involved for the spinning particle case, the extension should be feasible.

\item Extending our numerical and analytical codes to more complicated orbits, generic spin orientation, and Kerr spacetime are all important future steps. 
So far, the radiated fluxes have been computed for circular equatorial orbits in Kerr spacetime \cite{Harms:2015ixa,Chen:2019hac}, but the local force has not been calculated.
Making these calculations will further test our flux balance expression, which will be useful to ultimately drive fully generic orbit evolutions.

\item It is slightly unsatisfactory that the evolution equation for the spin tensor is dependent on the local metric perturbation. 
While this is purely the first-order non-spinning metric perturbation which will already be needed for self-force calculations, it would be aesthetically pleasing if the spin-forcing term could be fully related to the asymptotic amplitudes of the Teukolsky equation. 
Much like the Carter constant, we would expect that this relation would not explicitly be in terms of the asymptotic fluxes of energy and angular momentum.

\item As a consequence of our numerical methodology, the final results of both numerical codes contain spurious contributions which are non-linear in the spin. 
This necessitates an expensive fitting procedure to accurately extract the desired linear-in-$\s$ piece. 
In more complicated orbital and spin situations where each numerical computation is orders of magnitude more costly, this fitting may potentially be a significant problem. 
Thus, developing a code which can directly compute purely the linear-in-$\s$ contributions would be extremely valuable. 
Alternatively this may be a situation where high-order post-Newtonian expansions, which analytically extract the linear-in-$\s$ terms, may be a useful approach to cover large portions of the parameter space. 
Since the spin-dependent contributions are second order in the mass ratio, their accuracy requirements are significantly lower than those for the first-order fluxes, and thus the errors introduced by the post-Newtonian approximation will be substantially less important. 

\item The results of this work should be incorporated into practical inspiral evolution schemes. The conservative effects from a spinning secondary have been examined \cite{Burko:2015sqa,Warburton:2017sxk}, but as yet, the influence of the dissipative spin effects remains to be explored.

\item In this work we have concentrated on the dissipative sector, but one could also calculate conservative effects. 
These effects are not directly important for EMRI modelling as they contribute to the waveform phase at one order below the required post-adiabatic order. 
Nonetheless they are potentially very interesting when comparing with other approaches to the two-body problem. 
Calculation of conservative gauge invariants for a spinning secondary has been done in the PN regime for the redshift invariant \cite{Bini:2018zde}.
Extending this to numerical calculations in the strong-field and to other invariants \cite{Dolan:2013roa, Bini:2014ica, Dolan:2014pja, Bini:2014zxa, Nolan:2015vpa} is a natural next step.

\end{enumerate}

\acknowledgments
This work makes use of the Black Hole Perturbation Toolkit.
S.~A. acknowledges support by the EU H2020 under ERC Starting Grant, no.~BinGraSp-714626.
NW gratefully acknowledges support from a Royal Society - Science Foundation Ireland University Research Fellowship. We thank Adam Pound and Abraham Harte for helpful discussions, and thank Josh Mathews and Geoffrey Comp\`ere for comments on a draft of this work. S.D.~acknowledges financial support from the Science and Technology Facilities Council (STFC) under Grant No.~ST/P000800/1.
J.M. acknowledges support by the Sherman Fairchild Foundation and by NSF grants PHY-1708212 and PHY-1708213 at Caltech.

\appendix

\section{Additional details for the derivation of the source terms}

We now present explicit details of the computation of the sources for the Teukolsky equation.
As a representative example, consider Eq.~\eqref{eq:splus2_T0}, which is obtained by acting with $\edth\edth$ on $T_{ll}$ [given in Eq.~\eqref{eq:T-proj}].
Focusing on the $\delta_r\delta'_\phi$ term in $T_{ll}$, we have
\begin{equation}
\f{1}{2\pi} \int dt\, e^{i \omega t}\int {}_2\bar{Y}^{\ell m}(\theta,\phi)\,
\edth\edth \left(\f{K_2^{tr}}{f r^2\sin\theta} \delta_r \delta_\theta \delta'_\phi\right) \sin\theta\, d\theta d\phi. \label{eq:K2tr_term_in_FD}
\end{equation}
Applying \eqref{eq:int_by_parts} twice and shifting the derivative on $\delta_\phi$ onto the harmonic by integrating by parts, this becomes
\begin{equation}
\f{1}{2\pi} \int dt\, e^{i \omega t}\int i m \sqrt{(\ell-1)\ell(\ell+1)(\ell+2)}{}_0\bar{Y}^{\ell m}(\theta,\phi)\,
\left(\f{K_2^{tr}}{2 f r^4\sin\theta} \delta_r \delta_\theta \delta_\phi\right) \sin\theta\, d\theta d\phi.
\end{equation}
We can now immediately perform the integrals to obtain
\begin{equation}
i m \sqrt{(\ell-1)\ell(\ell+1)(\ell+2)} {}_0\bar{Y}^{\ell m}(\tfrac{\pi}{2},0)\,
\f{K_2^{tr}}{2 f r^4}
\end{equation}
along with the condition $\omega = m \Omega$. The remaining terms in Eq.~\eqref{eq:splus2_T0} can be
computed in a similar fashion, but starting with $\delta_\phi$ instead of $\delta'_\phi$, which 
results in an overall factor of $im$ for the latter. The expressions for the other terms in Eq.~\eqref{eq:splus2_T}
can be simplified in a similar fashion keeping in mind that the operator $\th$ contains partial
derivatives with respect to $t$ and $r$ coordinates, which introduces terms involving $\Omega$ and $r$.

\section{Variation of parameters weighting coefficients}\label{apdx:weighting_coeffs}

The variation-of-parameters weighting coefficients that appear in Sec.~\ref{sec:Teuk_source} are given by
\begin{align}
 {}_2 &C_{\ell m \omega} (r_\p) = \frac{2 \sqrt{(\ell -1) \ell  (\ell +1) (\ell +2)} \, {}_0 \bar{Y}_{\ell m}(\tfrac{\pi}{2},0)}{\Delta_\p^3 W(r_\p)} \bigg[-\sigma r_\p^2  f_\p^2 K_3^{tt} {}_2 R_{\ell m \omega}'(r_\p)
 \nonumber \\
 & \quad
  + \Big(r_\p^2 f_\p^2 K_{01}^{tt}-2 \sigma  r_\p f_\p (r_\p f'_\p+f_\p)K_3^{tt}-2 i \mathit{m} \sigma r_\p^2  f_\p K_2^{tr}+ \sigma r_\p^2 K_1^{rr}\Big) {}_2 R_{\ell m \omega}(r_\p) \bigg]
 \nonumber \\
  &-\frac{8 i \sqrt{(\ell -1) (\ell +2)} \, {}_1 \bar{Y}_{\ell m}(\tfrac{\pi}{2},0)}{\Delta_\p^3 W(r_\p)} \bigg[ \tfrac{1}{2} \sigma r_\p^4  f_\p^2 K_3^{t\phi} {}_2 R_{\ell m \omega}''(r_\p)
 \nonumber \\
 & \quad
  + \Big(-\tfrac{1}{2} r_\p^4 f_\p^2 K_{01}^{t\phi}+\tfrac{1}{2} \sigma r_\p^3  f_\p (4 r_\p f'_\p+6 f_\p+i\omega r_\p  )K_3^{t\phi} +\tfrac{1}{2} i \mathit{m} \sigma r_\p^4  f_\p K_2^{r\phi}\Big){}_2 R_{\ell m \omega}'(r_\p) 
  \nonumber \\
  & \quad
   +\Big(
   r_\p^3 ( r_\p f'_\p+f_\p+\tfrac{i\omega}{2} r_\p )(i \mathit{m} \sigma K_2^{r\phi}-f_\p K_{01}^{t\phi} )
   \nonumber \\
   & \qquad
   +\tfrac{1}{2} r_\p^2 \sigma  \big(r_\p^2 f'_\p (2 f'_\p+i \omega )+2 r_\p f_\p (r_\p f''_\p+6 f'_\p+2 i \omega )+6 f_\p^2\big)K_3^{t\phi} \Big){}_2 R_{\ell m \omega}(r_\p) \bigg]
   \nonumber \\
   &
   -\frac{2 \, {}_2 \bar{Y}_{\ell m}(\tfrac{\pi}{2},0)}{\Delta_\p^3 W(r_\p)} \bigg[
     \Big(r_\p^6 f_\p^2 K_{01}^{\phi \phi }+2 r_\p^5 \sigma  f_\p (-3 r_\p f'_\p-6 f_\p-i r_\p \omega )K_3^{\phi \phi } \Big){}_2 R_{\ell m \omega}''(r_\p) 
     - \sigma r_\p^6 f_\p^2 K_3^{\phi \phi } {}_2 R_{\ell m \omega}'''(r_\p)
   \nonumber \\
   & \quad
   + \Big(2 r_\p^5 f_\p (2 r_\p f'_\p+3 f_\p+i r_\p \omega )K_{01}^{\phi \phi } 
   \nonumber \\
   & \qquad
   -r_\p^4 \sigma  \big(r_\p^2 (5 i \omega  f'_\p+6 {f'_\p}^2-\omega ^2)+6 r_\p f_\p (r_\p f''_\p+8 f'_\p+3 i \omega )+34 f_\p^2\big)K_3^{\phi \phi }\Big){}_2 R_{\ell m \omega}'(r_\p)
   \nonumber \\
   & \quad
   +{}_2 R_{\ell m \omega}(r_\p) \Big(r_\p^4 \big(r_\p^2 (3 i \omega  f'_\p+2 {f'_\p}^2-\omega ^2)+2 r_\p f_\p (r_\p f''_\p+6 f'_\p+3 i \omega )+4 f_\p^2\big)K_{01}^{\phi \phi } 
   \nonumber \\
   & \qquad -r_\p^3 \sigma   \big(3 r_\p^2 (2 f'_\p+i \omega ) (r_\p f''_\p+4 f'_\p+2 i \omega )+2 r_\p f_\p (r_\p^2 f'''(r_\p)+12 r_\p f''_\p+34 f'_\p+15 i \omega )+16 f_\p^2\big)K_3^{\phi \phi }\Big)
   \bigg]
\end{align}
and
\begin{align}
 {}_{-2} &C_{\ell m \omega} (r_\p) = 
 \frac{\sqrt{(\ell -1) \ell  (\ell +1) (\ell +2)} {}_0 \bar{Y}_{\ell m}(\tfrac{\pi}{2},0)}{2\Delta_\p^{-1} W(r_\p) r_\p^3 f_\p^2}  \bigg[
   - \sigma r_\p f_\p^2 K_3^{tt} {}_{-2} R_{\ell m \omega}'(r_\p) 
  \nonumber \\ & \quad
   + \big(r_\p f_\p^2 K_{01}^{tt}+\sigma  r_\p K_1^{rr}+2 i \mathit{m} \sigma r_\p f_\p K_2^{tr}+2 \sigma f_\p^2 K_3^{tt}\big){}_{-2} R_{\ell m \omega}(r_\p) \bigg]
 \nonumber \\
 &
 +\frac{2 i \sqrt{(\ell -1) (\ell +2)} {}_{-1} \bar{Y}_{\ell m}(\tfrac{\pi}{2},0)}{\Delta_\p^{-1} W(r_\p)}
  \bigg[\tfrac{1}{2} \sigma  K_3^{t\phi} {}_{-2} R_{\ell m \omega}''(r_\p)
  \nonumber \\
  & \quad
  +\frac{1}{2 r_\p^2 f_\p}\Big(-r_\p^2 f_\p K_{01}^{t\phi} + \sigma   (-r_\p^2 f'_\p-2 r_\p f_\p+2 M-i r_\p^2 \omega )K_3^{t\phi}-i \mathit{m} \sigma r_\p^2 K_2^{r\phi}\Big){}_{-2} R_{\ell m \omega}'(r_\p) 
  \nonumber \\
  & \quad
  +\frac{1}{2 r_\p^3 f_\p^2}\Big(r_\p f_\p(r_\p^2 f'_\p+2 r_\p f_\p-2 M+i r_\p^2 \omega )K_{01}^{t\phi} -\mathit{m} \sigma r_\p (-i r_\p^2 f'_\p-2 i r_\p f_\p+2 i M+r_\p^2 \omega )K_2^{r\phi} 
  \nonumber \\ & \qquad
  +\sigma  \big(-f_\p (r_\p^3 f''_\p+4 M)+r_\p f'_\p (r_\p^2 f'_\p-2 M+i r_\p^2 \omega )+2 r_\p f_\p^2\big)K_3^{t\phi} \Big){}_{-2} R_{\ell m \omega}(r_\p) 
  \bigg]
 \nonumber \\
 &
 + \frac{{}_{-1} \bar{Y}_{\ell m}(\tfrac{\pi}{2},0)}{2\Delta_\p^{-1} W(r_\p)}
 \bigg[-\frac{1}{f_\p}\Big(r_\p^2 f_\p K_{01}^{\phi \phi }+\sigma  (r_\p^2 f'_\p-2 M+2 i r_\p^2 \omega )K_3^{\phi \phi } \Big){}_{-2} R_{\ell m \omega}''(r_\p) +\sigma r_\p^2  K_3^{\phi \phi } {}_{-2} R_{\ell m \omega}'''(r_\p)
 \nonumber \\ & \quad
 -\frac{1}{r_\p f_\p^2}\Big(\sigma  \big(2 f_\p (r_\p^3 f''_\p+M+i r_\p^2 \omega )+4 r_\p f'_\p (M-i r_\p^2 \omega )-2 r_\p^3 {f'_\p}^2+3 r_\p^2 f_\p f'_\p+2 r_\p f_\p^2
 \nonumber \\ & \qquad
 +r_\p \omega  (r_\p^2 \omega +2 i M)\big)K_3^{\phi \phi } -r_\p f_\p (r_\p^2 f'_\p+2 r_\p f_\p-2 M+2 i r_\p^2 \omega )K_{01}^{\phi \phi } \Big){}_{-2} R_{\ell m \omega}'(r_\p) 
 \nonumber \\ & \quad 
 -\frac{1}{r_\p^2 f_\p^3}\Big(\sigma \big[-2 r_\p^2 f'_\p (\omega -i f'_\p) (-i r_\p^2 f'_\p+2 i M+r_\p^2 \omega )+r_\p f_\p (2 r_\p (r_\p \omega ^2+f''_\p (M-i r_\p^2 \omega ))-3 r_\p^2 {f'_\p}^2
 \nonumber \\ & \qquad
 -f'_\p (3 r_\p^3 f''_\p+2 M+2 i r_\p^2 \omega ))+f_\p^2 (r_\p^4 f'''(r_\p)+3 r_\p^3 f''_\p+r_\p^2 f'_\p-2 M-2 i r_\p^2 \omega )\big]K_3^{\phi \phi }
 \nonumber \\ & \qquad
 - r_\p f_\p \big(r_\p (\omega -i f'_\p) (-i r_\p^2 f'_\p+2 i M+r_\p^2 \omega )+f_\p (r_\p^3 f''_\p+r_\p^2 f'_\p+2 M-2 i r_\p^2 \omega )\big)K_{01}^{\phi \phi } \Big){}_{-2} R_{\ell m \omega}(r_\p) 
 \bigg],
\end{align}
with $\Delta_0^{s+1} W(r_0)$ the invariant Wronskian.

\bibliographystyle{apsrev4-1}
\bibliography{spin_diss_references}

\end{document}